\documentclass[lettersize,journal]{IEEEtran}
\usepackage{amsmath,amsfonts}
\usepackage{array}

\usepackage{textcomp}
\usepackage{stfloats}
\usepackage{url}
\usepackage{verbatim}
\usepackage{graphicx}

\usepackage{hyperref}

\usepackage{amssymb}
\usepackage{booktabs}
\usepackage{algorithm}
\usepackage{multirow}
\usepackage{makecell}

\usepackage{subfigure}

\usepackage{color}
\usepackage{algorithm}
\usepackage{algpseudocode}

\renewcommand{\arraystretch}{1.1}

\newcommand{\eg}{\textit{e}.\textit{g}.}
\newcommand{\ie}{\textit{i}.\textit{e}.}
\newcommand{\etal}{\textit{et al}.}

\usepackage{forloop}
\newcounter{ct}

\def\BibTeX{{\rm B\kern-.05em{\sc i\kern-.025em b}\kern-.08em
    T\kern-.1667em\lower.7ex\hbox{E}\kern-.125emX}}
\usepackage{balance}

\begin{document}
\title{Image Enhancement Based on Pigment Representation}
\author{Se-Ho Lee, Keunsoo Ko, and Seung-Wook Kim
\thanks{This work was supported by the Research Fund of The Catholic University of Korea, in 2024.}
\thanks{Se-Ho Lee is with the Department of Computer Science and Artificial Intelligence/Center for Advanced Image Information Technology, Jeonbuk National University, Jeonju 54896, South Korea (e-mail: seholee@jbnu.ac.kr)}
\thanks{Keunsoo Ko is with the Department of Artificial Intelligence, The Catholic University of Korea, Bucheon 14662, South Korea. (e-mail: ksko@catholic.ac.kr).}
\thanks{Seung-Wook Kim is with the Division of Electronic and Communication Engineering, Pukyong National University, Busan 48513, South Korea (e-mail: swkim@pknu.ac.kr)}
\thanks{Corresponding author: Keunsoo Ko (e-mail: ksko@catholic.ac.kr).}
\thanks{The source codes are available at \href{https://github.com/jbnu-vilab/pigment_enhancement}{https://github.com/jbnu-vilab/pigment\_enhancement}}
\thanks{This paper has been accepted for publication in IEEE Transactions on Multimedia (TMM), 2025.}}

\markboth{Journal of \LaTeX\ Class Files,~Vol.~18, No.~9, September~2020}%
{How to Use the IEEEtran \LaTeX \ Templates}

\maketitle

\begin{abstract}
This paper presents a novel and efficient image enhancement method based on pigment representation. Unlike conventional methods where the color transformation is restricted to pre-defined color spaces like RGB, our method dynamically adapts to input content by transforming RGB colors into a high-dimensional feature space referred to as \textit{pigments}. The proposed pigment representation offers adaptability and expressiveness, achieving superior image enhancement performance. The proposed method involves transforming input RGB colors into high-dimensional pigments, which are then reprojected individually and blended to refine and aggregate the information of the colors in pigment spaces. Those pigments are then transformed back into RGB colors to generate an enhanced output image. The transformation and reprojection parameters are derived from the visual encoder which adaptively estimates such parameters based on the content in the input image. Extensive experimental results demonstrate the superior performance of the proposed method over state-of-the-art methods in image enhancement tasks, including image retouching and tone mapping, while maintaining relatively low computational complexity and small model size.
\end{abstract}


\section{Introduction}
\label{sec:introduction}
\begin{figure*}[]
    \centering
    {
        {\includegraphics[width=\textwidth]{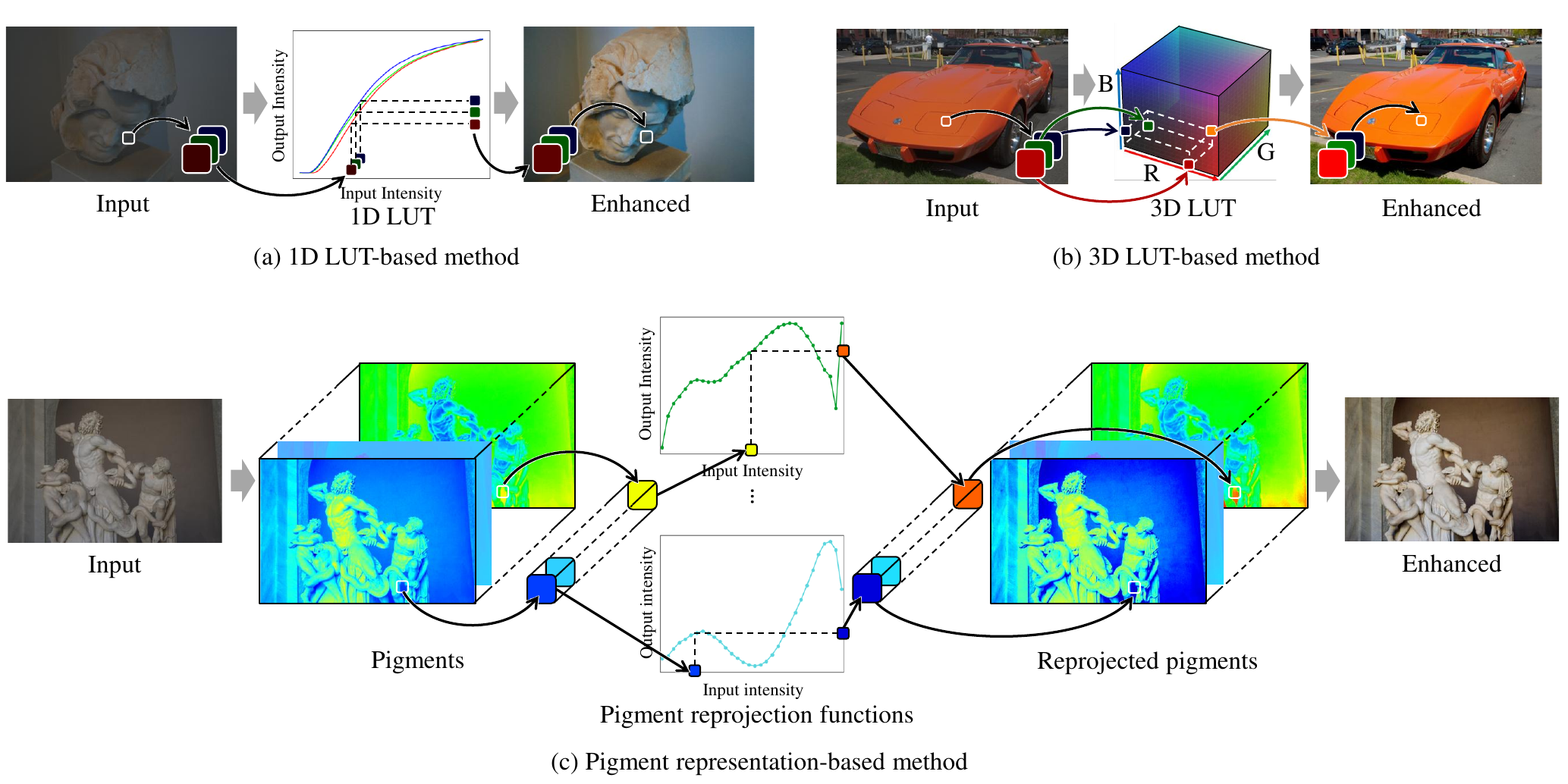}}

    }
    \caption
    {
        Comparisons of the (a) 1D LUT-based method, (b) 3D LUT-based method, and (c) the proposed pigment representation-based method. In (c), the input image is first converted into a set of pigments, which are then transformed using pigment reprojection functions. The reprojected pigments are subsequently combined to reconstruct the enhanced image.
    }
    \vspace{-0.1cm}
    \label{fig:comparison}
\end{figure*}
Nowadays, people take digital photographs casually with various mobile devices. However, due to their limited size, the quality of photos captured by mobile devices is poorer than that of professional-grade cameras, especially in uncontrolled environments such as inadequate lighting, back-lighting, nighttime, haze, or non-uniform illumination. Image enhancement can alleviate this problem. For example, professional applications like Photoshop and Lightroom provide various tools to enhance image quality effectively. However, using these applications requires a lot of effort, even for professionals, since it demands a high level of manual adjustment skill and user experience.

To resolve such inconveniences, many studies have been conducted to perform image enhancement automatically. They can be classified into two categories: dense mapping-based and global transformation-based methods. Dense mapping-based methods~\cite{FirstDeepLearning,HDRNet,DPE,DeepLPF,MAXIM,RSFNet,ZeroDCE,STARDCE,LLNet,DRBN,TMM3,TMM5,TMM6,TMM7,TMM8} learn dense end-to-end color mappings between the input and output image pairs via deep neural networks. These methods adaptively derive and apply the mapping function for each pixel according to the information in neighboring pixels. While dense mapping-based methods can achieve promising results, they usually suffer from heavy computational and memory burdens, limiting their practicalities. 
On the other hand, global transformation-based methods~\cite{gonzalez,HistEq1, HistEq2, HistEq3, HistEq4,Aesthetic,Reinforcement1,GlobalAndLocal,CSRNet,3DLUT,Adaint,SepLUT,LUT2,HashLUT,Reinforcement2,Reinforcement3} improve the image quality by applying global transformation functions, which are usually controlled by a set of parameters. Early efforts have predominantly concentrated on global transformation-based image enhancement methods, including gamma correction~\cite{gonzalez} and histogram-equalization methods~\cite{HistEq1, HistEq2, HistEq3, HistEq4}, due to their simplicity and consistent outcomes. However, these methods face limitations in modeling diverse color mappings between input and enhanced images because they rely on simplistic image priors like the uniform distribution of a target histogram. Consequently, they may yield sub-optimal outcomes with exaggerated artifacts or noise. 

To overcome this limitation, global transformation-based methods with learnable parameters have been proposed~\cite{Aesthetic,Reinforcement1,GlobalAndLocal,CSRNet,Reinforcement2,Reinforcement3}. These methods utilize a series of transformations, where individual color channel values or other factors such as contrast and saturation are separately mapped based on 1D lookup tables~(LUTs). However, despite their computational efficiency, these methods struggle to comprehensively capture complex color mappings between input and enhanced images. This limitation arises from their incapacity to model interactions among different components, as each transformation operates independently. For more complex color mapping, 3D LUT-based methods have been introduced~\cite{3DLUT,Adaint,SepLUT,LUT2,HashLUT,4DLUT,CoTF}, where each axis of a 3D LUT represents the intensity values for one of three color channels. The 3D LUTs enable sophisticated component-correlated transforms that take into account correlations among different color channels. However, they still exhibit limitations in achieving content-specific adaptability, as the transformations are performed within a pre-defined RGB color space and thus may not effectively reflect the semantic or contextual characteristics of the input. In addition, they suffer from a substantial memory footprint and increased training complexity. Moreover, when using a larger LUT size, there is a risk of inadequate cell utilization due to the curse of dimensionality. Typically, colors in a single input image only cover a small portion of the entire color space, resulting in limited cell utilization. Consequently, a significant portion of the model becomes unnecessary.
For instance, our experiments confirm that \cite{3DLUT} utilizes an average of only 4.95\% of 3D LUT cells for each image in the MIT-Adobe5K dataset~\cite{Adobe5K}.

The key limitations in existing approaches can be summarized as follows: 1) 1D LUT-based methods often suffer from limited expressiveness; 2) 3D LUT-based methods face the curse of dimensionality and inefficient cell utilization; 3) both approaches rely heavily on fixed color spaces, whose coordinate systems often lack the flexibility to model the complex relationship inherent in perceptual and physical color interactions, such as pigment mixing in painting~\cite{tan2018pigmento,sochorova2021practical}.

Our motivation lies in overcoming these limitations. To this end, we propose an image enhancement approach that increases expressiveness by mapping RGB colors into a high-dimensional, input-dependent color space customized for each image. This adaptive representation is designed to capture richer and more flexible color relationships beyond fixed coordinate systems. Additionally, we apply one-dimensional transformations within this space to avoid the curse of dimensionality.

Specifically, Fig.~\ref{fig:comparison} compares the 1D LUT-based method, 3D LUT-based method, and the proposed method. Both 1D LUT-based~(Fig.~\ref{fig:comparison}(a)) and 3D LUT-based~(Fig.~\ref{fig:comparison}(b)) methods rely solely on pre-defined color spaces, such as the RGB color space~\cite{GlobalAndLocal,3DLUT,Adaint,SepLUT,LUT2}, the CIE LAB color space~\cite{Aesthetic}, other factors like contrast, saturation, color, etc.~\cite{Reinforcement1}, or CNN features~\cite{CSRNet}. Therefore, these methods may not effectively account for the color distribution of the content in a target image; on the other hand, the proposed method customizes the conversion, which expands RGB colors into a higher-dimensional feature space and refines the expanded features, referred to as \textit{pigment} representation, as illustrated in Fig.~\ref{fig:comparison}(c). To obtain each pigment, we apply a simple linear transformation to the RGB colors of input images. The parameters for this transformation are not fixed values but rather estimated by the visual encoder, where the visual encoder takes the entire input image as its input. Consequently, the proposed pigment representation achieves adaptability to the colors of the input content, rather than depending on predetermined configurations of color spaces.

Also, to avoid the curse of dimensionality, the proposed method employs one-dimensional reprojection functions as done in 1D LUT-based methods. Unlike 3D LUT-based methods, where a significant portion of the model capacity remains unutilized due to the limited coverage of colors in input images, the proposed method ensures that a substantial portion is actively utilized by employing one-dimensional reprojection functions. On average, the proposed method utilizes 90.09\% of 1D LUT cells for each image in the MIT-Adobe5K dataset~\cite{Adobe5K}, which is 18.2 times higher than that of the 3D LUT-based method~\cite{Adobe5K}. Furthermore, in contrast to 1D LUT-based methods, the proposed pigment representation enables complex color mapping even with one-dimensional reprojection functions. Since each pigment is obtained by a linear transformation of input RGB colors, each pigment reprojection function can modify input colors along the correlated color direction, which can be seen as the component-correlated transform. Subsequently, by aggregating information from abundant reprojected pigments, we can achieve a high level of expressiveness in color mapping. Therefore, the proposed pigment representation effectively addresses model capacity issues while preserving the benefits of utilizing one-dimensional reprojection functions.

The proposed method consists of five stages: visual encoder, pigment expansion, pigment reprojection, pigment blending, and RGB reconstruction. In the first stage, the visual encoder predicts key parameters for image enhancement: pigment expansion weights, reprojection offsets, and RGB reconstruction weights. Following this, the pigment expansion stage performs the conversion of the RGB color space in an input image into a set of pigments. The process of color space conversion involves calculating the weighted sum of the RGB colors. Instead of employing fixed weight values, we utilize the pigment expansion weights obtained from the visual encoder as weight values. This dynamic approach ensures that the pigments are derived considering the content of the input image.
Moving on to the third stage, pigment reprojection reprojects the input pigments using a set of pigment reprojection functions. Each pigment reprojection function consists of a set of input/output pigment values determined by the reprojection offsets from the visual encoder. Through the reprojection of numerous pigments, a set of 1D reprojection functions can infer complex mappings of input/target images.
In the fourth stage, we enhance pigment correlations to enable the representation of more complex color mappings by employing pigment blending with two convolution layers. In the final stage, the RGB reconstruction acquires reconstructed images by calculating the weighted sum of the blended pigments. We utilize the RGB reconstruction weights from the visual encoder as the weight values. 

The main contributions of this paper can be summarized as follows:
\begin{itemize}
\item We propose a novel image enhancement method that includes pigment expansion, which involves customized conversions of RGB colors into pigments. Leveraging the predicted weights from the visual encoder, pigment expansion effectively adapts to the input content.
\item We introduce pigment reprojection, a technique that reprojects a set of pigments using corresponding one-dimensional reprojection functions. Leveraging the abundance of pigments, the proposed method achieves high expressiveness in color mapping.
\item The proposed method demonstrates superior performance in tasks such as photo retouching and tone mapping, as evidenced by its performance on the MIT-Adobe FiveK~\cite{Adobe5K} and PPR10K~\cite{PPR10K} datasets, respectively. Notably, it achieves this high-performance level with relatively low computational complexity and a compact model size.
\end{itemize}

\section{Related Work}
\label{sec:related}

The objectives of image enhancement are closely related but different between color enhancement, contrast enhancement~\cite{lee2013contrast,TCSVT_CE}, dehazing~\cite{kim2013optimized,TCSVT_dehazing}, and bit-depth enhancement~\cite{wan2016image,TCSVT_depth}. This section briefly reviews color enhancement techniques.

\subsection{Dense Mapping-Based Image Enhancement}
With the advance of deep neural networks, various dense mapping-based approaches have been proposed by learning the dense color mapping functions between input and output image pairs.
Yan~\etal~\cite{FirstDeepLearning} introduced the first deep learning-based method for estimating a pixel-wise color mapping function using handcrafted feature descriptors. Gharbi~\etal~\cite{HDRNet} proposed a method that uses bilateral filtering to enhance images, incorporating local and global adjustments. Local features predict local affine transformation coefficients via bilateral grid processing, while global features capture overall image characteristics. Chen~\etal~\cite{DPE} utilized the U-Net architecture~\cite{UNet} with global features, allowing for local adjustments at the pixel level while considering high-level global information. Moran~\etal~\cite{DeepLPF} introduced a novel method for automatic image enhancement using spatially local filters, including Elliptical, Graduated, and Polynomial filters. They estimated the coefficients of these filters and applied them to enhance images. Tu~\etal~\cite{MAXIM} employed a U-Net-shaped network with a multi-axis MLP-based architecture for various image processing tasks, including enhancement, deblurring, and denoising. Ouyang~\etal~\cite{RSFNet} proposed a method that simultaneously generates filter arguments, such as saturation, contrast, hue, and attention maps for regions associated with each filter. They utilized linear summations on filtered images to obtain enhanced images, facilitating a more extensive range of filter classes that can be trained more efficiently. He~\etal~\cite{TCSVT1} introduced a U-Net-based multi-scale adjustment network for both global illumination and local contrast. Furthermore, they incorporated a wavelet-based attention network to efficiently detect and reduce noise in the frequency domain, providing particular benefits for improving images.

\begin{figure*}[]
    \centering
    {
        {\includegraphics[width=\textwidth]{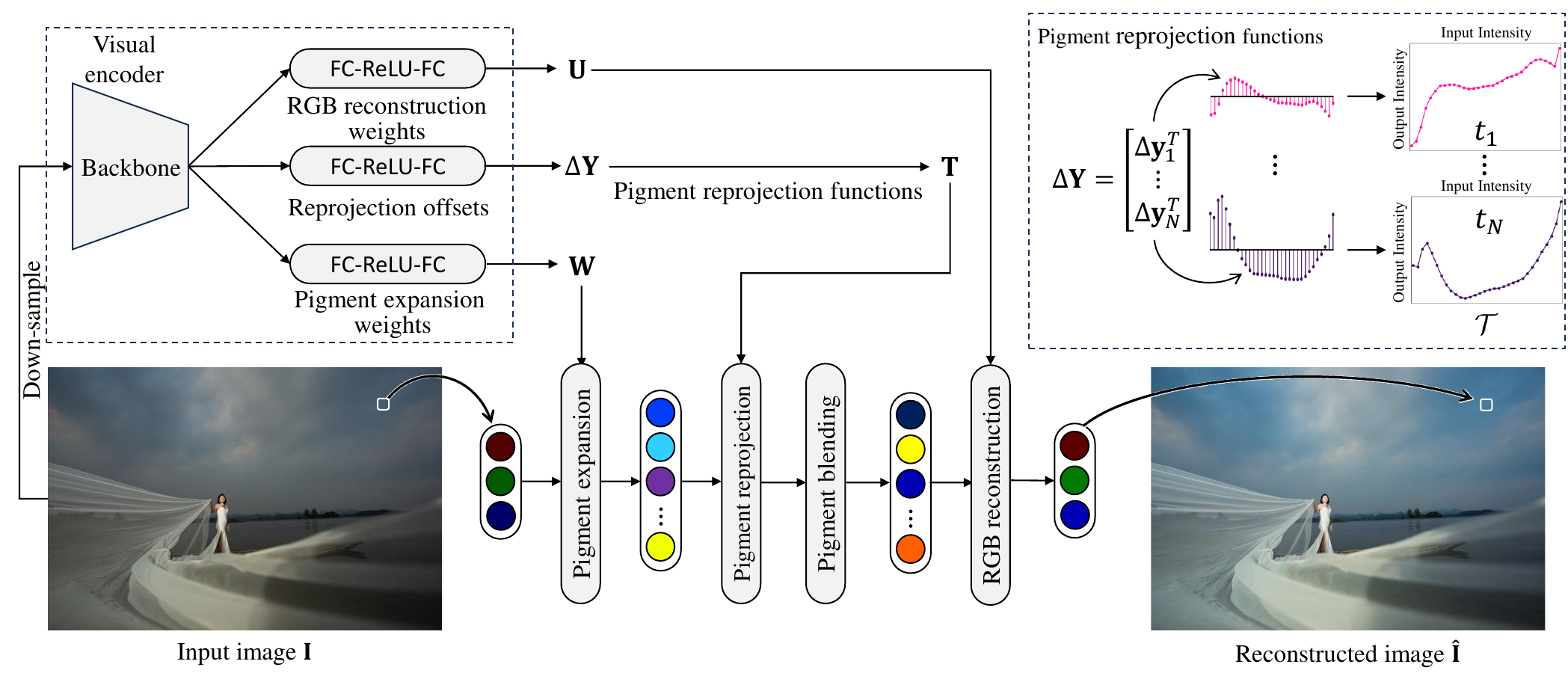}}

    }
    \vspace{-0.3cm}
    \caption
    {
        An overall framework of the proposed method.
    }
    \vspace{-0.1cm}
    \label{fig:framework}
\end{figure*}

\subsection{Global Transformation-Based Image Enhancement}
In contrast, global transformation-based image enhancement improves image quality by applying global transformation functions, typically determined by a set of parameters.
The representative global transformation-based image enhancement methods are based on gamma correction~\cite{gonzalez} and histogram-equalization~\cite{HistEq1, HistEq2, HistEq3, HistEq4}. The gamma correction utilizes a single parameter to map input intensities into output ones. Conversely, histogram-equalization-based approaches enhance image quality by adjusting the histogram of an input to match that of a target, providing computational simplicity.

Nevertheless, approaches like gamma correction and histogram-equalization methods are inadequate in representing complex color mappings between input and enhanced images.
Therefore, learnable global transformation-based methods have been proposed~\cite{Aesthetic,Reinforcement1,GlobalAndLocal,CSRNet}. Deng~\etal~\cite{Aesthetic} introduced a global transformation function utilizing a piece-wise transformation function in the CIE LAB color space. They divided each channel into three to five segments, each controlled by a single parameter. Hu~\etal~\cite{Reinforcement1} employed a series of global operations, \eg,~editing contrast, saturation, color, etc., learned through reinforcement learning. 
Kim~\etal~\cite{GlobalAndLocal} placed the global enhancement network and local enhancement network in series to enhance input images. While the U-Net~\cite{UNet}-based local enhancement network improves the image quality based on spatial filtering, the global enhancement network obtains channel-wise intensity transformation functions for each RGB channel separately. 
He~\etal~\cite{CSRNet} introduced a model consisting of a base network and a conditional network. The base network processes each pixel independently, while the conditional network extracts overall features to create a condition vector. Input images are enhanced through global feature modulation, which modulates the features from the base network by adjusting parameters based on the condition vector.

However, the methods mentioned earlier~\cite{Aesthetic,Reinforcement1,GlobalAndLocal,CSRNet}, which can be seen as 1D LUT-based methods, still have limitations in modeling capability.
To address the limitation, 3D LUT-based methods have recently been introduced~\cite{3DLUT,Adaint,SepLUT,LUT2,HashLUT,4DLUT,CoTF}. Zeng~\etal~\cite{3DLUT} researched learning the scene-adaptive 3D LUTs, wherein the 3D color transformation for an input image is stored within a 3D lattice. This approach conducts a uniform sampling of input color values and then stores the corresponding output color values in the 3D lattice. To determine the output color values for each input pixel, the method identifies the nearest sampling points within the sampled 3D lattice and subsequently employs trilinear interpolation. However, the uniform sampling approach used in this 3D LUT method may impose limitations on its expressiveness. Addressing this concern, Yang~\etal~\cite{Adaint} proposed an alternative method that learns non-uniform sampling intervals within the 3D color space. This innovation allows for a more flexible allocation of sampling points, offering improved adaptability in the color transformation process. 
Also, Wang~\etal~\cite{LUT2} and Liu~\etal~\cite{4DLUT} obtain pixel-wise category maps to improve the robustness in local regions for the traditional 3D LUT. While Wang~\etal~\cite{LUT2} introduced the concept of learnable spatial-aware 3D LUTs, emphasizing adaptability to spatial features, Liu~\etal~\cite{4DLUT} extended 3D LUTs into 4D LUTs by adding a category-aware dimension. Yang~\etal~\cite{SepLUT} introduced a separable image-adaptive lookup table-based method. Their approach involves applying independent color transformation to each RGB color channel using 1D LUT and subsequently employing 3D LUT-based color transformation. This dual-LUT strategy ensures mutual support, where the 3D LUT enhances the capability to blend color components, and the 1D LUTs redistribute input colors to maximize the cell utilization of the 3D LUT. Also, Zhang~\etal~\cite{HashLUT} introduced an efficient hash form of 3D LUT to reduce the number of parameters. Li~\etal~\cite{CoTF} integrated global and pixel-wise transformations with a relation-aware modulation module. The global transformation uses 3D LUTs for overall contrast and detail, while the pixel transformation adjusts local context.

\section{Proposed Method}
\label{sec:proposed}
Fig.~\ref{fig:framework} illustrates the overall framework of the proposed method. First, the visual encoder estimates the pigment expansion weights, pigment reprojection offsets, and RGB reconstruction weights adaptively according to an input image. Subsequently, the proposed method follows a sequential process involving pigment expansion, pigment reprojection, pigment blending, and RGB reconstruction. 
The key idea of the proposed color enhancement method is efficient color conversion and refinement in the high-dimensional feature space, referred to as pigment representation.
The following subsections present the proposed method in detail.

\subsection{Visual Encoder}
\label{ssec:backbone}
The visual encoder generates essential parameters for pigment expansion, pigment reprojection, and RGB reconstruction stages, called pigment expansion weights, reprojection offsets, and RGB reconstruction weights, respectively. The proposed visual encoder obtains these parameters by utilizing the entire input image as an input, playing a crucial role in achieving adaptiveness to the image content. Specifically, the input image is down-sampled to $256\times256$ and fed into a backbone network, which outputs 512-channel features for the input images. The details of the backbone network are described in Section~\ref{ssec:implementation}. Each prediction for pigment expansion weights, reprojection offsets, and RGB reconstruction weights uses the 512-channel features as inputs and involves the use of two fully-connected layers with a ReLU activation in between. The first fully-connected layer for each prediction extracts $C$ channel features, where $C=128$ is set to adopt a bottleneck architecture, reducing the number of parameters compared to 512. Subsequently, the final fully-connected layers predict pigment expansion weights, reprojection offsets, and RGB reconstruction weights, respectively.

\subsection{Pigment Expansion}
\label{ssec:pigment_expansion}
Whereas previous global transformation-based methods perform their transformations within pre-defined color spaces such as RGB~\cite{GlobalAndLocal,3DLUT,Adaint,SepLUT,LUT2} or nonlinear representations including the CIE LAB color space~\cite{Aesthetic}, handcrafted attributes like contrast, saturation, and color~\cite{Reinforcement1}, or CNN features~\cite{CSRNet}, these approaches are inherently limited in adaptability due to their reliance on fixed transformation schemes. In contrast, we transform colors through a customized, content-aware conversion from the RGB space into a high-dimensional pigment space, referred to as pigment expansion.
It yields several key advantages: 1) adapting to the content of an input image and 2) enabling complex color mapping for image enhancement through the use of abundant pigments.

Let $\mathbf{c}(i)=\left[c^{\mathrm{(r)}}(i),c^{\mathrm{(g)}}(i),c^{\mathrm{(b)}}(i)\right]^T$ represent the intensity for red (r), green (g), and blue (b) color channels at pixel location $i$, where $c^{(k)}(i) \in [0,1]$ for ${k}\in\{\mathrm{r},\mathrm{g},\mathrm{b}\}$. For brevity, we omit the pixel location $i$ and thus denote $\mathbf{c}(i)$ as $\mathbf{c}$. %
To construct a pigment representation, we utilize pigment expansion weights $\mathbf{W}=[\mathbf{w}_1,...,\mathbf{w}_N] \in \mathbb{R}^{3 \times N}$, where $\mathbf{w}_{n}=[{w}_n^{(\mathrm{r})}, {w}_n^{(\mathrm{g})}, {w}_n^{(\mathrm{b})} ]^T$, for $n=1,...,N$. Here, $\mathbf{w}_{n}$ is utilized to customize the transformation of the input RGB color $\mathbf{c}$ into the $n$-th pigment $p_n$, so applying $\mathbf{W}$ can transform the color $\mathbf{c}$ into abundant pigments. Noticing that the pigment expansion weights $\mathbf{W}$ are derived via the visual encoder, they allow the construction of a high-dimensional feature space that adapts to the content of the input image. To constrain the range of each pigment to $[0,1]$, we normalize each column vector of $\mathbf{W}$ to obtain $\hat{\mathbf{W}}= [\hat{\mathbf{w}}_1,...,\hat{\mathbf{w}}_N] $, ensuring every component of $\hat{\mathbf{w}}_n$ lies within the range of $[0,1]$, and the sum of $\hat{\mathbf{w}}_n$ equals $1$. Thus, $\hat{\mathbf{w}}_n$ is given by
\begin{equation}
\label{eq:wn}
\hat{\mathbf{w}}_n=\frac{\sigma(\mathbf{w}_n)}{\sum_{k} \sigma({w}_n^{(k) })},
\end{equation}
where $\sigma (\cdot)$ denotes the sigmoid function.
Then, we can obtain the pigment representation $\mathbf{p} = {[p_1,...,p_N]}^T$ by performing a linear transformation of $\mathbf{c}$ as
\begin{equation}
\label{eq:f}
    \mathbf{p} = \hat{\mathbf{W}}^{T}\mathbf{c}.
\end{equation}

\begin{figure}[t]
    \centering
    {
        {\includegraphics[width=8cm]{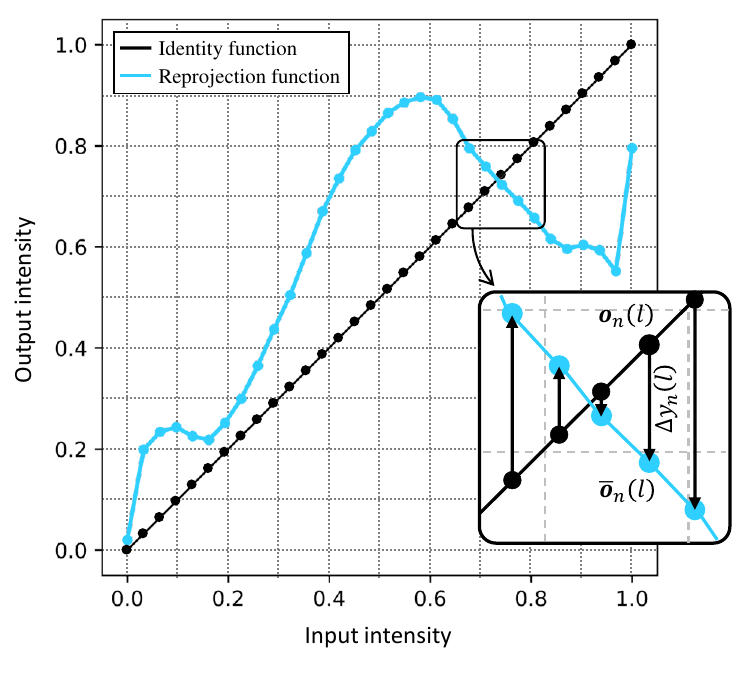}}

    }
    \vspace{-0.3cm}
    \caption
    {
        An example of pigment reprojection function. 
    }
    \vspace{-0.1cm}
    \label{fig:DCP}
\end{figure}

\subsection{Pigment Reprojection}
\label{ssec:pigment_reproj}
To reproject pigments in a way that is favorable for representing the enhanced image, we generate a pigment reprojection function $t_n$ for each $n$-th pigment.
Fig.~\ref{fig:DCP} outlines the process of obtaining a pigment reprojection function $t_n$. For every $n$-th pigment, we individually generate a pigment reprojection function $t_n$ to reproject each pigment in a way that is favorable for representing the enhanced image.

To derive the pigment reprojection functions, we uniformly sample $L$ points from the identity function depicted as dots on the black line in Fig.~\ref{fig:DCP}, with each point denoted as $\mathbf{o}_n(l)$:
\begin{equation}
	\label{eq:on}
	\mathbf{o}_n(l) = \left(x_{n}(l), y_{n}(l)\right),
\end{equation}
where $l \in \{1,2,\dots,L\}$, and $x_{n}(l)$ and $y_{n}(l)$ are given as $\frac{l-1}{L-1}$. 
Then, we obtain the tweaked point $\bar{\mathbf{o}}_n(l)$, which are depicted as dots on the blue line in Fig.~\ref{fig:DCP}, as 
\begin{align}
\label{eq:baro}
\bar{\mathbf{o}}_n &= (x_n(l), \bar{y}_n(l)) \nonumber \\
                   &= (x_n(l), y_n(l)+ \Delta y_n(l)),
\end{align}
where $\Delta y_n(l)$ is the reprojection offset.
The proposed method utilizes the visual encoder to extract $N \times L$ reprojection offsets, allowing the pigment reprojection to vary with the contents of input images. These offsets are represented as
\begin{equation}
	\label{eq:deltaY}
	\Delta \mathcal{Y} = \{\Delta y_n(l)|n=1,...,N, l=1,...,L\}.
\end{equation}
Note that each adjacent tweaked points $\bar{\mathbf{o}}_n(l)$ and $\bar{\mathbf{o}}_n(l+1)$ are connected linearly.
For the pigment $p_{n}$ within the range of $[x_n(l),x_n(l+1)]$, the pigment reprojection function $t_n$ maps $p_{n}$ to $\bar{p}_{n}$ by employing linear interpolation between $\bar{y}_n(l)$ and $\bar{y}_n(l+1)$, as given by
\begin{align}
\label{eq:barfn}
\bar{p}_n &= t_{n}(p_n) \nonumber \\
          &= \alpha_n^{(l)} \cdot  \bar{y}_n(l) + (1-\alpha_n^{(l)}) \cdot \bar{y}_n(l+1), 
\end{align}
where $\alpha_n^{(l)}=(x_n(l+1) - p_n)/(x_n(l+1) - x_n(l))$, which takes on a value of $1$ when $p_n$ reaches to $x_n(l)$ and $0$ when $p_n$ reaches $x_n(l+1)$. Therefore, using Eqs. (\ref{eq:on})-(\ref{eq:barfn}), the reprojected pigments $\bar{\mathbf{p}} = {[ \bar{p}_{1},...,\bar{p}_{N} ]}^T$ can be obtained from the input pigments $\mathbf{p} = {[ p_{1},...,p_{N} ]}^T$ for each pixel. Since each pigment $p_n$ represents the linear transformation of input RGB color $\mathbf{c}$, reprojecting each pigment $p_n$ to $\bar{p}_n$ enables component-correlated transforms that consider correlations among different color channels. Additionally, by performing pigment reprojections on abundant pigments, we can obtain complex color mappings for image enhancement.

\subsection{Pigment Blending}
\label{ssec:pigment_blending}
Before RGB reconstruction, we aggregate and refine the reprojected pigments in the high-dimensional feature space to enhance the expressiveness of the proposed method. We incorporate a stage with an efficient neural network model. This model consists of two $1\times1$ convolution layers that generate $N$-channel features, followed by batch normalization~\cite{BN} and ReLU activation. We refer to this stage as pigment blending, and its output $\hat{\mathbf{p}}$ is obtained by passing $\bar{\mathbf{p}}$ through this model.
In this stage, independently reprojected pigments from the pigment reprojection stage are effectively combined, enhancing the non-linearity of the process to strengthen the expressiveness of the proposed method.

\subsection{RGB Reconstruction}
\label{ssec:rgb_reconstruction}
Subsequently, we reconstruct an RGB image from the enhanced pigments $\hat{\mathbf{p}}$ using a linear transformation similar to the pigment expansion.
Specifically, we obtain the reconstruction weights 
$\mathbf{U}=[\mathbf{u}_1,...,\mathbf{u}_N] \in \mathbb{R}^{3 \times N}$, where $\mathbf{u}_{n}=[u_n^{(\mathrm{r})}, u_n^{(\mathrm{g})}, u_n^{(\mathrm{b})} ]^T$, using the visual encoder. The reconstruction weights $\mathbf{U}$ are used to perform the customized conversion of pigments into RGB colors. The reconstructed color $\mathbf{\hat{c}}=[\hat{c}^{(\textrm{r})},\hat{c}^{(\textrm{g})},\hat{c}^{(\textrm{b})}]^T$ is then determined by
\begin{equation}
\label{eq:base_point4}
    \mathbf{\hat{c}}=\mathbf{U}\mathbf{\hat{p}}.
\end{equation}
Finally, each reconstructed color $\mathbf{\hat{c}}$ constitutes the reconstructed image $\hat{\mathbf{I}}$.

\subsection{Loss Function}
\label{ssec:Loss}
Let $\hat{\mathbf{I}}$ and $\tilde{\mathbf{I}}$ represent the reconstructed image and its corresponding ground-truth image, respectively.
The loss function between $\hat{\mathbf{I}}$ and $\tilde{\mathbf{I}}$ is defined as 
\begin{equation}
	\label{eq:loss}
	\ell = \| \hat{\mathbf{I}} - \tilde{\mathbf{I}} \|_1  +\lambda\sum_{i=2,4,6}\| \phi^{(i)} (\hat{\mathbf{I}}) - \phi^{(i)} (\tilde{\mathbf{I}}) \|_1,
\end{equation}
where the first term computes the average L1 distance between the RGB colors of the reconstructed and ground-truth images, while the second term, known as perceptual loss~\cite{Perceptual}, measures the absolute average error between the CNN features of the reconstructed and ground-truth images. Here, we utilize VGG-16~\cite{VGG}, pretrained on ImageNet~\cite{ImageNet}, and $\phi^{(i)}$ refers to the output feature of the $i$-th layer in VGG-16. The parameter $\lambda$ is empirically set to $0.1$ to balance between the two loss terms.

\section{Experimental Results}
\label{sec:exp}
In this section, we show the efficacy of the proposed method by comparing it with state-of-the-art methods on the MIT-Adobe FiveK~\cite{Adobe5K} and PPR10K~\cite{PPR10K} datasets. Moreover, we conduct analyses and ablation studies to confirm the efficacy of the proposed components. For all comparisons, when quantitative performance metrics are available, we use the reported performance values from the state-of-the-art methods as stated in their corresponding papers. Otherwise, we utilize source codes provided by authors with default settings.

\def\arraystretch{1.2}
\begin{table}[t]
    \centering
    \caption{Quantitative comparisons on the MIT-Adobe FiveK dataset~(480p)~\cite{Adobe5K} for the photo retouching application. Results marked with ``-'' indicate that they are not available due to incompatibility with our hardware settings or the unavailability of the corresponding method's source code. The best result is boldfaced.}
    \label{table:adobe5K}
    \resizebox{1\columnwidth}{!}
    {
        \small
        \begin{tabular}{c | ccccc} 
            \Xhline{3\arrayrulewidth}
            Methods & PSNR & SSIM & $\Delta E_{ab}$ & \# of Params. & Runtime \\
            \Xhline{3\arrayrulewidth}
            UPE~\cite{UPE} & 21.88 & 0.853 & 10.80 & 927.1K & - \\
            DPE~\cite{DPE} & 23.75 & 0.908 & 9.34 & 3.4M & - \\
            HDRNet~\cite{HDRNet} & 24.66 & 0.915 & 8.06 & 483.1K & - \\
            CSRNet~\cite{CSRNet} & 25.17 & 0.921 & 7.75 & 36.4K & 0.71ms \\
            DeepLPF~\cite{DeepLPF} & 24.73 & 0.916 & 7.99 & 1.7M & 36.69ms \\
            3D LUT~\cite{3DLUT} & 25.29 & 0.920 & 7.55 & 593.5K & 0.80ms \\
            SepLUT~\cite{SepLUT} & 25.47 & 0.921 & 7.54 & 119.8K & 6.20ms \\
            AdaInt~\cite{Adaint} & 25.49 & 0.926 & 7.47 & 619.7K & 1.89ms \\
   RSFNet~\cite{RSFNet} & 25.49 & 0.924 & 7.23 & 16.1M & 7.28ms \\
   4D LUT~\cite{4DLUT} & 25.50 & 0.931 & 7.27 & 924.4K & 1.25ms \\

   HashLUT~\cite{HashLUT} & 25.50 & 0.926 & 7.46 & 114.0K & - \\
   CoTF~\cite{CoTF} & 25.54 & 0.938 & 7.46 & 310.0K & 4.28ms \\
   Proposed & \textbf{25.82} & \textbf{0.939} & \textbf{7.15} & 765.0K & 1.43ms \\
            \Xhline{3\arrayrulewidth}
        \end{tabular}
    }
\end{table}

\subsection{Datasets}
\label{ssec:dataset}
\subsubsection{MIT-Adobe FiveK~\cite{Adobe5K}}
\label{sssec:fivek}
It contains 5,000 RAW images captured from various DSLR cameras and is commonly used for the photo retouching task. Each image is retouched by five experts~(A/B/C/D/E), with expert C's version used as the ground-truth, following \cite{3DLUT,Adaint,SepLUT}. The dataset is split into 4,500 and 500 image pairs for training and testing, respectively. We conduct experiments on the original 4K images and their 480p versions with down-sampling applied to the short side.

We assess the proposed method on two tasks: photo retouching and tone mapping. Both tasks use target images in the standard 8-bit sRGB format but differ in that photo retouching uses 8-bit sRGB inputs to enhance images, while tone mapping takes 16-bit CIE XYZ inputs to convert high-dynamic-range images into low-dynamic-range ones for standard displays with limited dynamic ranges.

For quantitative assessment, we employ the peak signal-to-noise ratio~(PSNR), structural similarity index measure~(SSIM)~\cite{SSIM}, and $\Delta E_{ab}$ metrics, which measure pixel-wise similarity, structural similarity, and the $L_{2}$-distance in the CIE LAB color space between the enhanced and ground-truth images, respectively. Higher PSNR and SSIM values indicate better performance, while a lower $\Delta E_{ab}$ value signifies better performance as well.

\begin{figure*}[t]
    \centering

    \subfigure{\includegraphics[width=4cm]{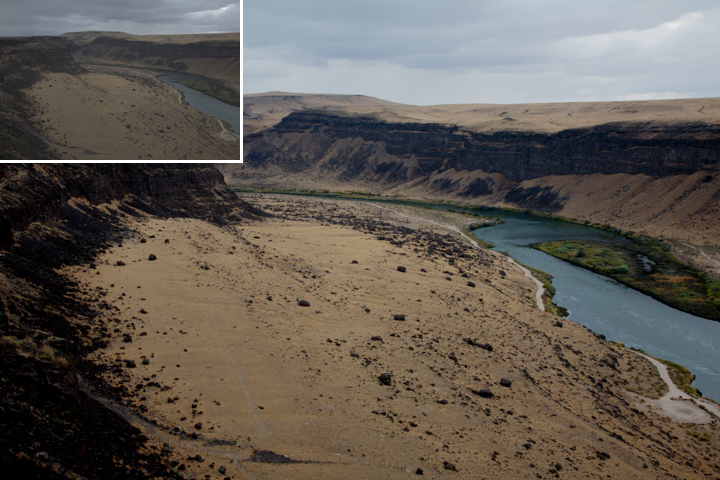}}\hspace{0.25cm}%
    \subfigure{\includegraphics[width=4cm]{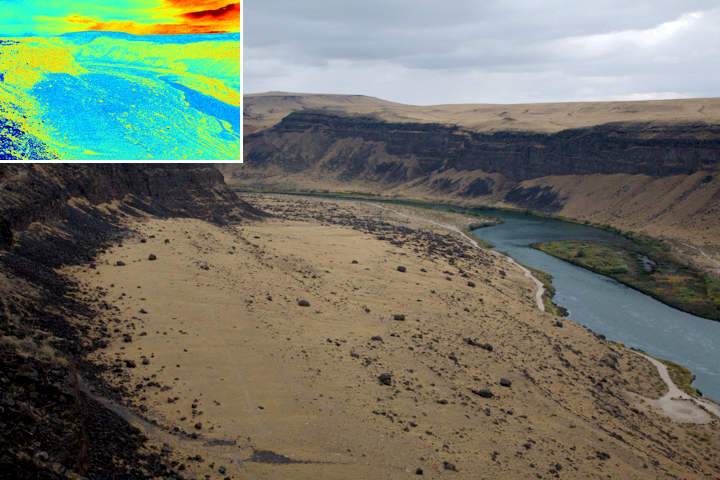}}\hspace{0.25cm}%
    \subfigure{\includegraphics[width=4cm]{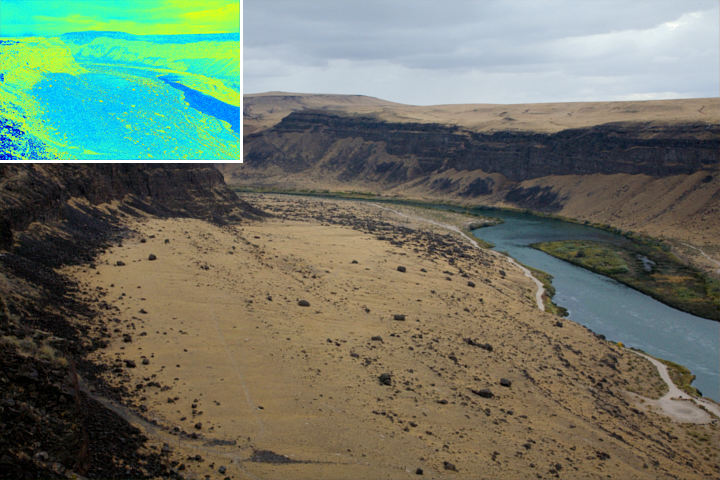}}\hspace{0.25cm}%
    \subfigure{\includegraphics[width=4cm]{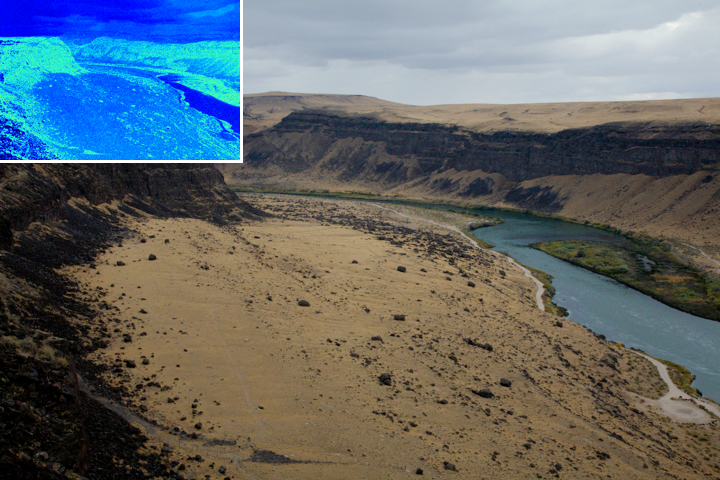}}\hspace{0.01cm}%
    \subfigure{\includegraphics[width=0.258cm]{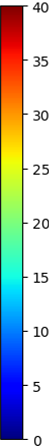}} \\[4pt]
    \vspace{-0.2cm}
    \subfigure{\includegraphics[width=4cm]{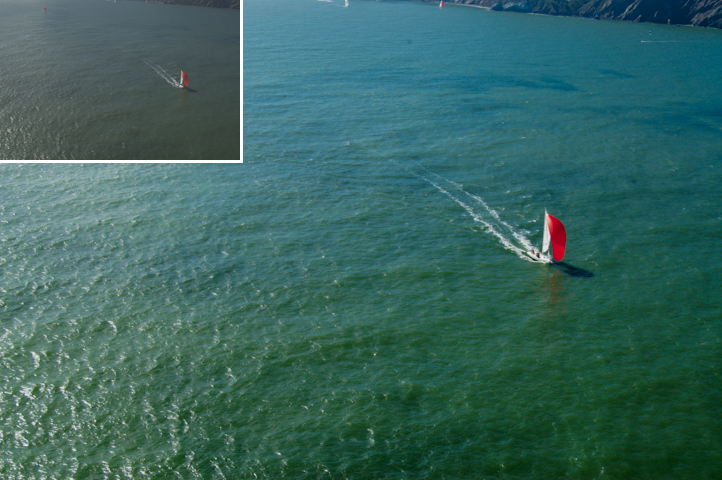}}\hspace{0.25cm}%
    \subfigure{\includegraphics[width=4cm]{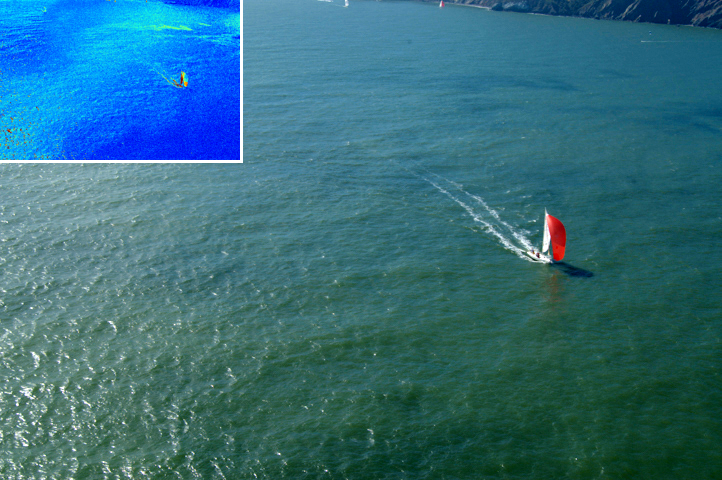}}\hspace{0.25cm}%
    \subfigure{\includegraphics[width=4cm]{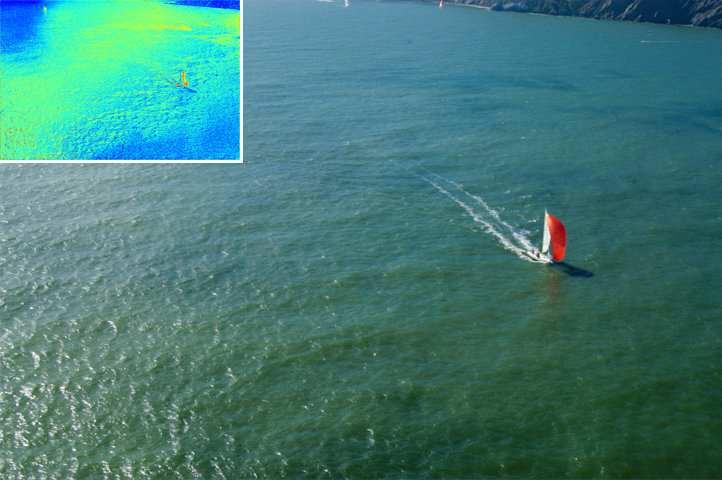}}\hspace{0.25cm}%
    \subfigure{\includegraphics[width=4cm]{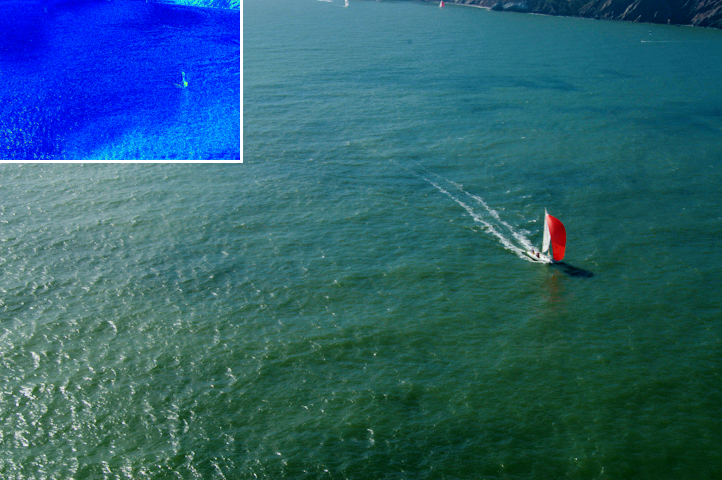}}\hspace{0.01cm}%
    \subfigure{\includegraphics[width=0.258cm]{Figures/res_h128_2/colorbar.png}} \\[4pt]
    \vspace{-0.2cm}
    \subfigure{\includegraphics[width=4cm]{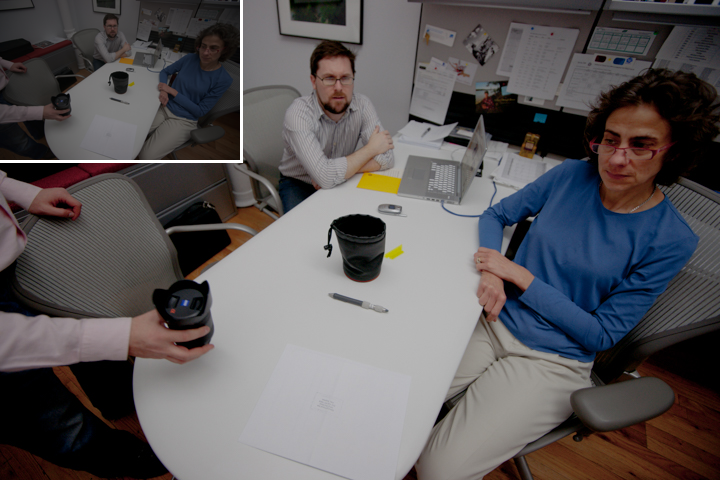}}\hspace{0.25cm}%
    \subfigure{\includegraphics[width=4cm]{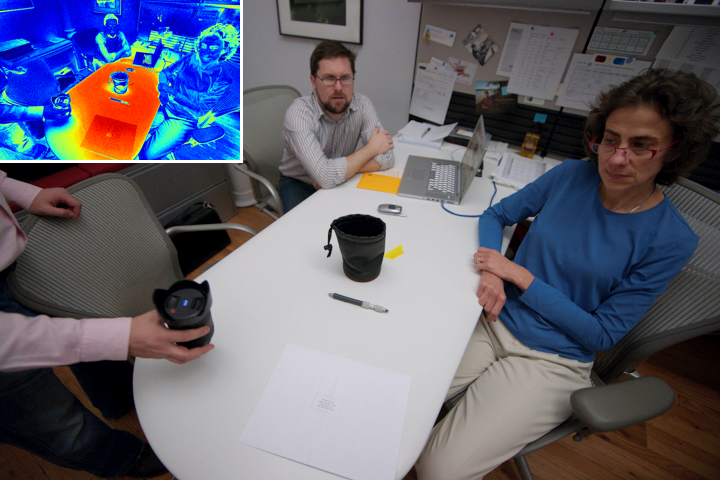}}\hspace{0.25cm}%
    \subfigure{\includegraphics[width=4cm]{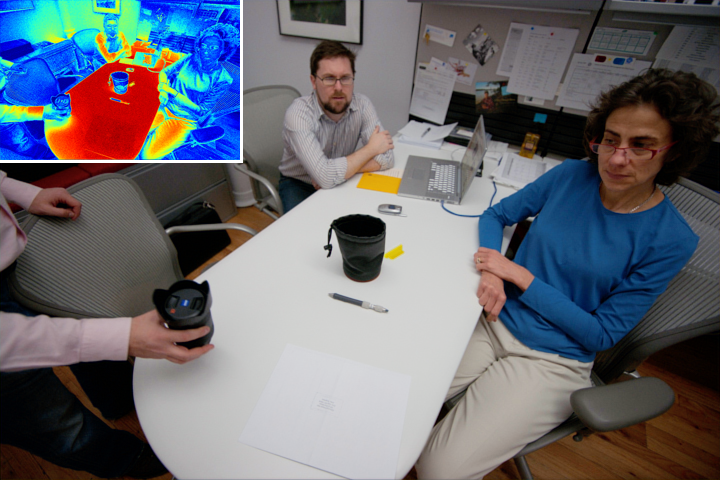}}\hspace{0.25cm}%
    \subfigure{\includegraphics[width=4cm]{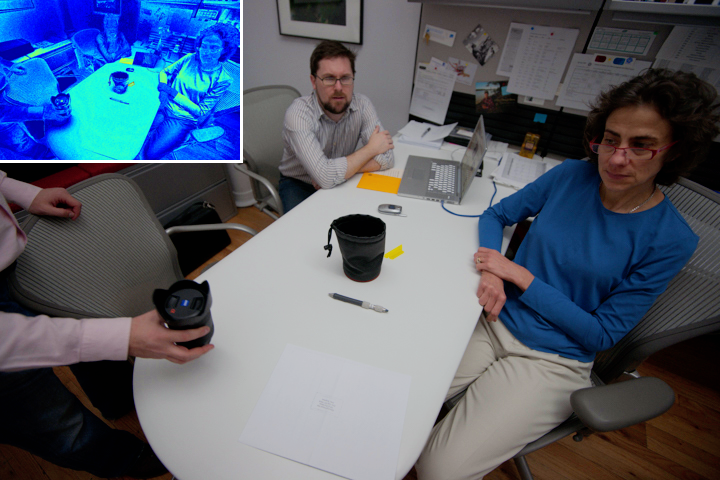}}\hspace{0.01cm}%
    \subfigure{\includegraphics[width=0.258cm]{Figures/res_h128_2/colorbar.png}} \\[4pt]
    \vspace{-0.2cm}
    \subfigure{\includegraphics[width=4cm]{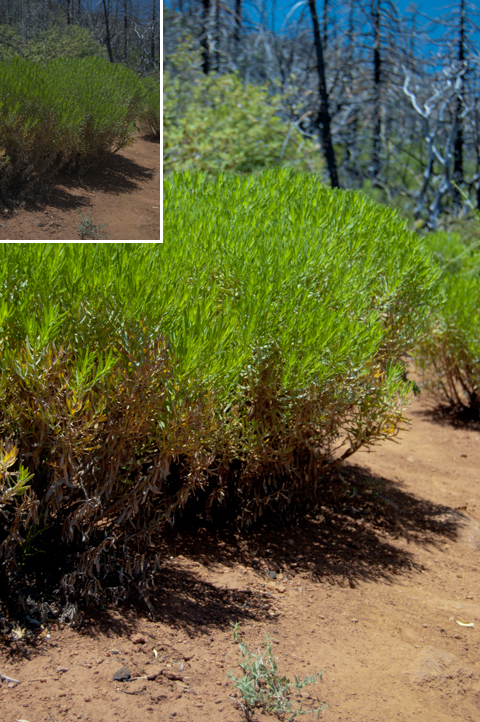}}\hspace{0.25cm}%
    \subfigure{\includegraphics[width=4cm]{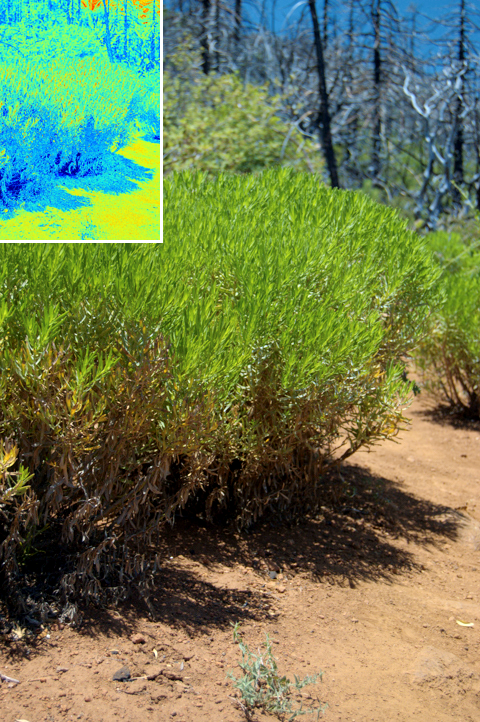}}\hspace{0.25cm}%
    \subfigure{\includegraphics[width=4cm]{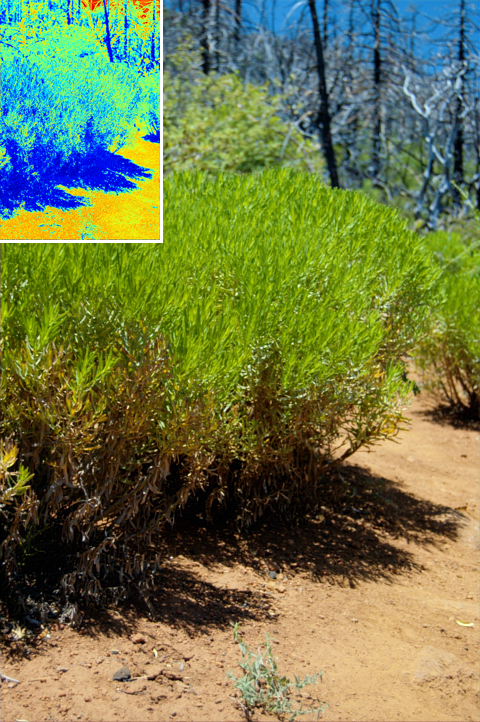}}\hspace{0.25cm}%
    \subfigure{\includegraphics[width=4cm]{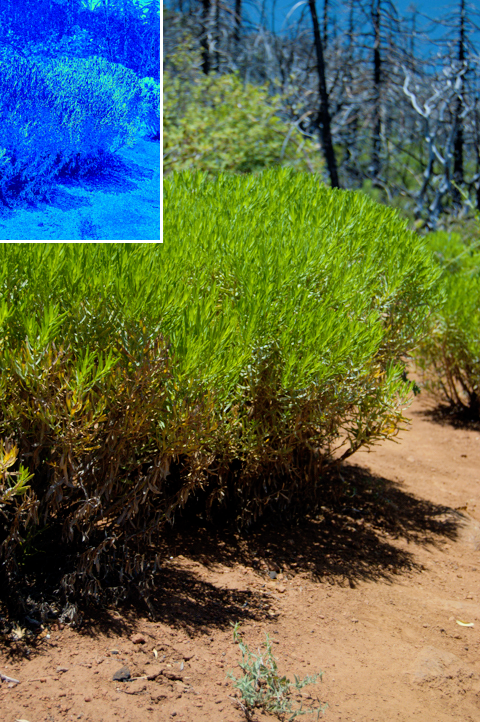}}\hspace{0.01cm}%
    \subfigure{\includegraphics[width=0.258cm]{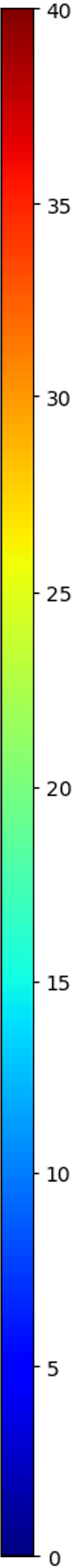}} \\[2pt]

    \makebox[4cm]{\small (a) GT}\hspace{0.25cm}%
    \makebox[4cm]{\small (b) 4D LUT}\hspace{0.25cm}%
    \makebox[4cm]{\small (c) CoTF}\hspace{0.25cm}%
    \makebox[4cm]{\small (d) Proposed}\hspace{0.01cm}%
    \makebox[0.258cm]{}%

    \vspace{-0.1cm}
    \caption{Qualitative comparisons on the MIT-Adobe FiveK dataset~\cite{Adobe5K} for image retouching: 
    (a) shows GT~(retouched by expert C) with its corresponding input image. 
    (b), (c), and (d) show the resultant images and their corresponding error maps obtained by 4D LUT~\cite{4DLUT}, 
    CoTF~\cite{CoTF}, and the proposed method, respectively.}
    \label{fig:AdobeComparison}
\end{figure*}

\subsubsection{PPR10K~\cite{PPR10K}}
\label{sssec:ppr10k}
It is also considered a photo retouching dataset and comprises 11,161 high-quality RAW portrait photos. Three distinct experiments can be conducted with three retouched versions~(a/b/c). As in \cite{PPR10K}, we partitioned the dataset into 8,875 and 2,286 pairs for training and testing, respectively. During training, we use images augmented by the dataset creator~\cite{PPR10K} as inputs. For the evaluation, we utilize the 360p version of the dataset and employ not only PSNR, SSIM, and $\Delta E_{ab}$ but also human-centered metrics~\cite{PPR10K}, namely PSNR$^{HC}$ and $\Delta E_{ab}^{HC}$, which specifically assess the quality of human subject reconstruction.

\subsection{Implementation Details}
\label{ssec:implementation}
We conduct both training and testing on a single NVIDIA 1080 GPU, implementing our model using PyTorch. For both the MIT-Adobe FiveK~\cite{Adobe5K} and PPR10K~\cite{PPR10K} datasets, training involves 400 epochs with a batch size of 16. The Adam optimizer~\cite{Adam} is utilized with a weight decay of $1\times10^{-5}$. A learning rate is initially set to $1\times10^{-4}$ and is adjusted using cosine learning rate decay. During training, we randomly crop input images to sizes of $256\times256$ and $512\times512$ for the MIT-Adobe FiveK and PPR10K datasets, respectively. Additionally, we perform random horizontal and vertical flipping for data augmentation. For testing, no cropping or flipping is applied to ensure consistent results. We empirically use $N=64$ and $L=32$ unless otherwise noted.

The backbone network in the visual encoder, as discussed in Section~\ref{ssec:backbone}, follows the configurations outlined in \cite{3DLUT,PPR10K} for a fair comparison. Specifically, we adopt the 5-layer backbone introduced in \cite{3DLUT} for the MIT-Adobe FiveK dataset~\cite{Adobe5K}. Conversely, for the PPR10K dataset~\cite{PPR10K}, we use ResNet-18~\cite{ResNet} pretrained on ImageNet~\cite{ImageNet}. 

\begin{figure*}
    \centering
    \subfigure{\includegraphics[width=4cm]{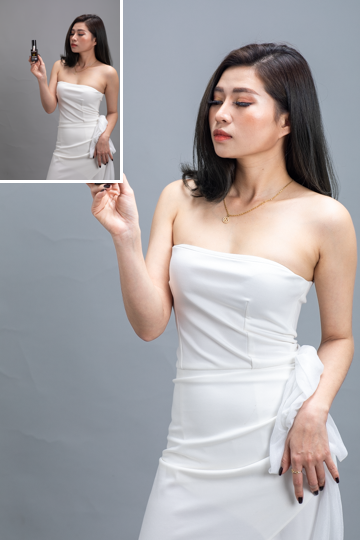}}  \hspace{0.02cm}
    \subfigure{\includegraphics[width=4cm]{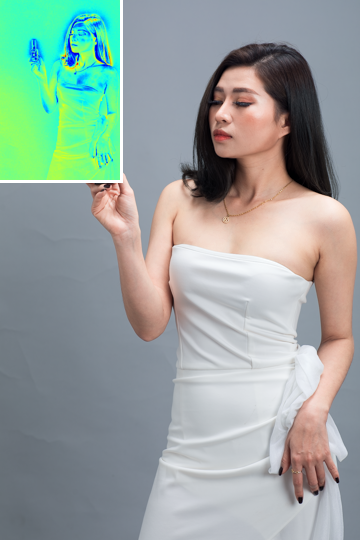}} \hspace{0.02cm}
    \subfigure{\includegraphics[width=4cm]{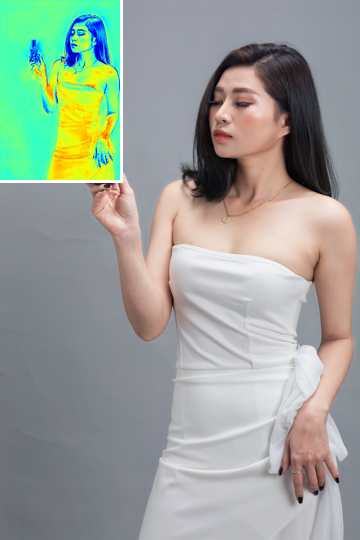}} \hspace{0.02cm}
    \subfigure{\includegraphics[width=4cm]{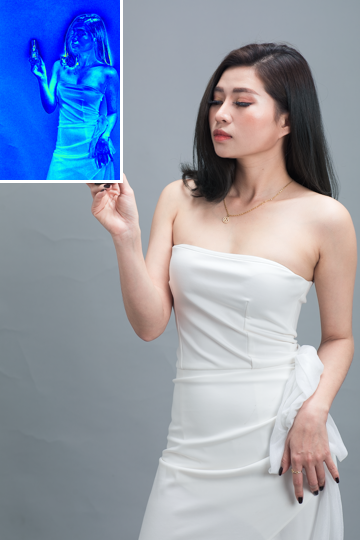}} \hspace{0.01cm}
    \subfigure{\includegraphics[width=0.258cm]{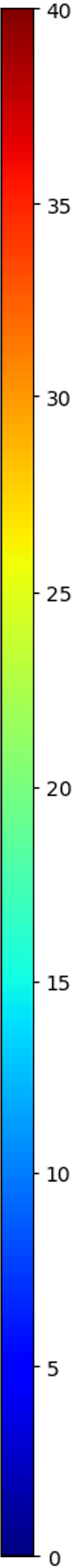}} \\
    \subfigure{\includegraphics[width=4cm]{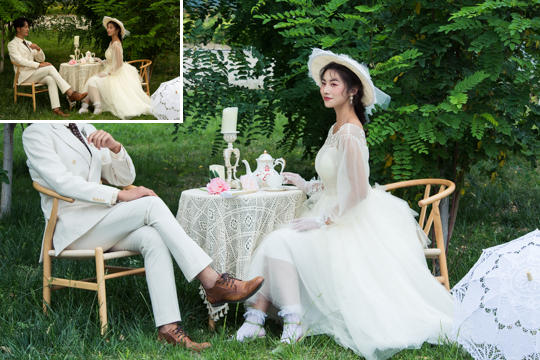}}  \hspace{0.02cm}
    \subfigure{\includegraphics[width=4cm]{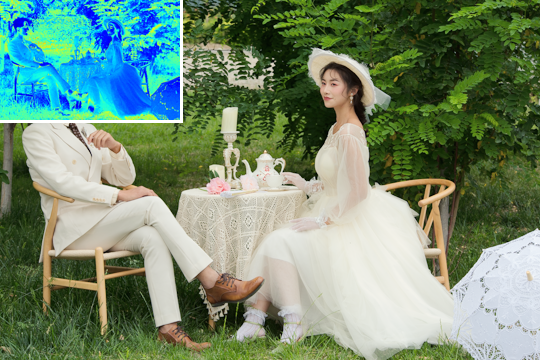}} \hspace{0.02cm}
    \subfigure{\includegraphics[width=4cm]{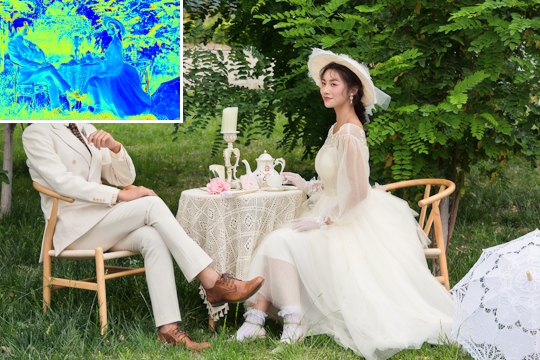}} \hspace{0.02cm}
    \subfigure{\includegraphics[width=4cm]{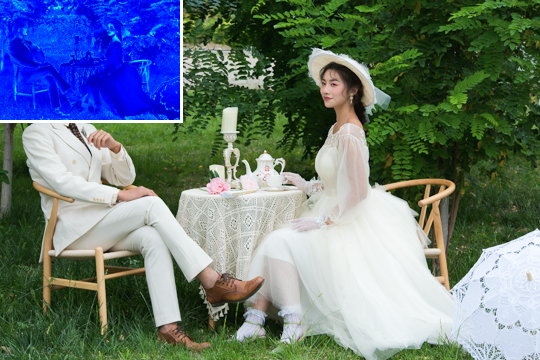}} \hspace{0.01cm}
    \subfigure{\includegraphics[width=0.258cm]{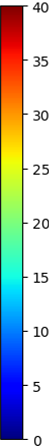}} \\
    \setcounter{subfigure}{0}
    \subfigure[GT]{\includegraphics[width=4cm]{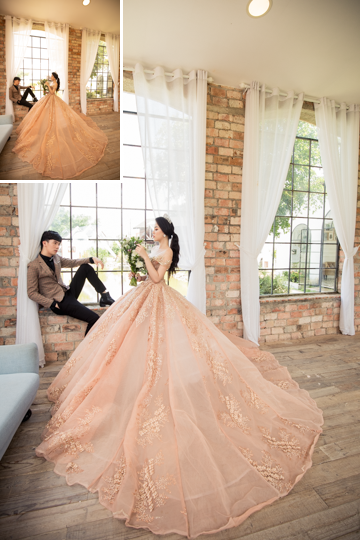}}  \hspace{0.02cm}
    \subfigure[AdaInt]{\includegraphics[width=4cm]{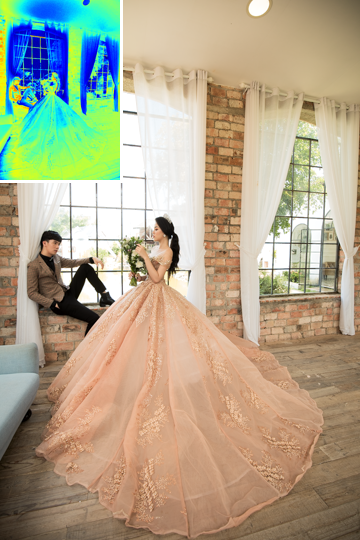}} \hspace{0.02cm}
    \subfigure[RSFNet]{\includegraphics[width=4cm]{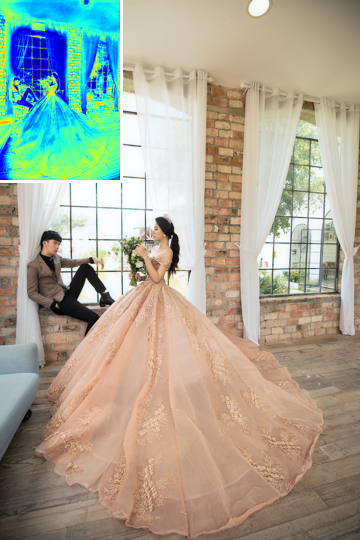}} \hspace{0.02cm}
    \subfigure[Proposed]{\includegraphics[width=4cm]{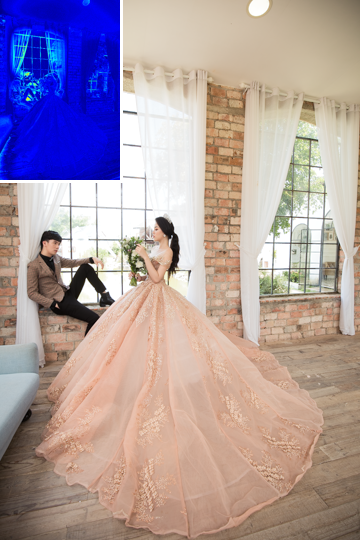}} \hspace{0.01cm}
    \subfigure{\includegraphics[width=0.258cm]{Figures/ppr10k/colorbar_long.png}}  \\
    \caption{Qualitative comparisons on the PPR10K dataset~\cite{PPR10K} for image retouching: (a) shows GT~(retouched by expert a) with its corresponding input image. (b), (c), and (d) show the resultant images and their corresponding error maps obtained by AdaInt~\cite{Adaint}, RSFNet~\cite{RSFNet}, and the proposed method, respectively.}
\label{fig:PPR10KComparison}
\vspace{-0.3cm}
\end{figure*}

\def\arraystretch{1.2}
\begin{table}[t]
    \centering
    \caption{Quantitative comparisons on the MIT-Adobe FiveK dataset~(4K)~\cite{Adobe5K}. The best result is boldfaced.}
    \label{table:adobe5K_4K}
    \resizebox{0.63\columnwidth}{!}
    {
        \small
        \begin{tabular}{c | ccc} 
            \Xhline{3\arrayrulewidth}
            Methods & PSNR & SSIM & $\Delta E_{ab}$ \\
            \Xhline{3\arrayrulewidth}
            UPE~\cite{UPE} & 21.65 & 0.859 & 11.09 \\
            HDRNet~\cite{HDRNet} & 24.52 & 0.921 & 8.20  \\
            CSRNet~\cite{CSRNet} & 24.82 & 0.926 & 7.94 \\
            3D LUT~\cite{3DLUT} & 25.25 & 0.932 & 7.59  \\
            SepLUT~\cite{SepLUT} & 25.43 & 0.932 & 7.56  \\
            AdaInt~\cite{Adaint} & 25.48 & 0.934 & 7.45  \\
            4D LUT~\cite{4DLUT} & 25.37 & 0.942 & 7.29 \\
            CoTF~\cite{CoTF} & 25.54 & 0.944 & 7.29 \\
     
   Proposed & \textbf{25.64} & \textbf{0.945} & \textbf{7.27} \\
            \Xhline{3\arrayrulewidth}
        \end{tabular}
    }
    \vspace{-0.3cm}
\end{table}

\def\arraystretch{1.2}
\begin{table}[t]
    \centering
    \caption{Quantitative comparisons on the PPR10K dataset~\cite{PPR10K}, where ``E'' denotes ``Expert'', and
a/b/c are ground-truths from three experts. The best and the second-best results are boldfaced and underlined, respectively. ``-'' indicates that the corresponding source codes are unavailable.}
    \label{table:PPR10K}
    \resizebox{1.0\columnwidth}{!}
    {
        \small
        \begin{tabular}{c| cc  cccc} 
            \Xhline{3\arrayrulewidth}
            Methods  & E & PSNR & SSIM & $\Delta E_{ab}$ & PSNR$^{HC}$ & $\Delta E_{ab}^{HC}$ \\
            \Xhline{3\arrayrulewidth}
            HDRNet~\cite{HDRNet} & a & 23.93 & - & 8.70 & 27.21 & 5.65 \\
            CSRNet~\cite{CSRNet} & a & 22.72 & - & 9.75 & 25.90 & 6.33 \\
            3D LUT~\cite{3DLUT} & a & 25.64 & - & 6.97 & 28.89 & 4.53  \\
            $\text{3D LUT}_{\text{HRP}}$~\cite{PPR10K} & a & 25.99 & - & 6.76 & 28.29 & 4.38  \\
            SepLUT~\cite{SepLUT} & a & 26.28 & 0.956 & 6.59 & 29.53 & 4.29\\    
            AdaInt~\cite{Adaint} & a & 26.33 & 0.956 & 6.56 & 29.57 & 4.26 \\
            RSFNet~\cite{RSFNet} & a & 25.58 & 0.919 & 7.29 & 28.83 & 4.74 \\
            4D LUT~\cite{4DLUT} & a &24.32 & 0.933 & 8.40 & 27.54 & 5.48  \\
            HashLUT~\cite{HashLUT} & a & 26.34 & - & 6.56  & - & -  \\
            CoTF~\cite{CoTF} & a & 23.89 & 0.910  & 9.95 & 26.47 & 6.53  \\
            Proposed & a & \underline{26.44} & \textbf{0.958} & \underline{6.50}  & \underline{29.68} & \underline{4.24} \\
            Proposed-Res34 & a &  \textbf{26.57} & \textbf{0.958} & \textbf{6.43} & \textbf{29.82} & \textbf{4.18} \\
                \hline
            HDRNet~\cite{HDRNet} & b & 23.93 & - & 8.84 & 27.21 & 5.74 \\
            CSRNet~\cite{CSRNet} & b & 23.76 & - & 8.77 & 27.01 & 5.68 \\
            3D LUT~\cite{3DLUT} & b  & 24.70 & - & 7.71 & 27.99 & 4.99  \\
            $\text{3D LUT}_{\text{HRP}}$~\cite{PPR10K} & b  & 25.06 & - & 7.51 & 28.36 & 4.85  \\
            SepLUT~\cite{SepLUT} & b & 25.23 & 0.949 & 7.49 & 27.92 & 5.03 \\
            AdaInt~\cite{Adaint} & b & 25.40 & 0.949 & 7.33 & 28.65 & 4.75 \\
            RSFNet~\cite{RSFNet} & b & 24.81 & 0.916 & 7.93 & 28.05 & 5.14\\
            4D LUT~\cite{4DLUT} & b & 23.93 & 0.933 & 8.71 & 27.17 & 5.66  \\
            HashLUT~\cite{HashLUT} & b & 25.42 & - & 7.40 & - & - \\
            CoTF~\cite{CoTF} & b & 23.90 & 0.910 & 9.97 & 26.46 & 6.50  \\
            Proposed & b &  \underline{25.46} & \underline{0.954} & \underline{7.23} & \underline{28.68} & \underline{4.69} \\
            Proposed-Res34 & b &  \textbf{25.51} & \textbf{0.955} & \textbf{7.17} & \textbf{28.73} & \textbf{4.65} \\
                \hline
            HDRNet~\cite{HDRNet} & c & 24.08 & - & 8.87 & 27.32 & 5.76 \\
            CSRNet~\cite{CSRNet} & c & 23.17 & - & 9.45 & 26.47 & 6.12 \\
            3D LUT~\cite{3DLUT} & c  & 25.18 & - & 7.58 & 28.49 & 4.92  \\
            $\text{3D LUT}_{\text{HRP}}$~\cite{PPR10K} & c  & 25.46 & - & 7.43 & 28.80 & 4.82  \\
            SepLUT~\cite{SepLUT} & c & 25.59 & 0.944 & 7.51 & 28.88 & 4.84 \\
            AdaInt~\cite{Adaint} & c & 25.68 & 0.943 & 7.31 & 28.93 & 4.76 \\
            RSFNet~\cite{RSFNet} & c & 25.52 & 0.913 & 7.52 & 28.85 & 4.87\\
            4D LUT~\cite{4DLUT} & c & 24.38 & 0.925 & 8.48 & 27.60 &  5.52  \\
            HashLUT~\cite{HashLUT} & c & 25.65 & - & 7.30  & - & -  \\
            CoTF~\cite{CoTF} & c & 24.26 & 0.903 & 10.02 & 26.82 & 6.55  \\
            Proposed & c &  \underline{25.78} & \underline{0.945} & \underline{7.24} & \underline{29.03} & \underline{4.70} \\
            Proposed-Res34 & c &  \textbf{25.83} & \textbf{0.946}  & \textbf{7.20} & \textbf{29.08} & \textbf{4.68} \\
                \hline
            \Xhline{3\arrayrulewidth}
        \end{tabular}
    }
\end{table}

\subsection{Comparison with State-of-the-Arts}
\label{ssec:SOTA}

\subsubsection{Photo retouching}
We compare the proposed method with twelve state-of-the-art photo retouching methods on the MIT-Adobe FiveK dataset~\cite{Adobe5K} at the resolution of 480p. As shown in Table~\ref{table:adobe5K}, the proposed method achieves superior performance in terms of PSNR, SSIM, and $\Delta E_{ab}$, showing a notable improvement of at least 0.28 in PSNR and 0.08 in $\Delta E_{ab}$. While the SSIM score is similar to CoTF~\cite{CoTF} with a margin of 0.001, it achieves significantly better PSNR and $\Delta E_{ab}$ by 0.28 and 0.31, respectively. These results confirm the proposed method's effectiveness in photo retouching by successfully capturing complex input-to-output color mappings using pigment representation. In terms of efficiency, the model has 765K parameters and executes in an average of 1.43ms on 480p images, which is 14.4\% slower but uses 17.2\% fewer parameters compared to 4D LUT~\cite{4DLUT}, while achieving superior performance across all metrics.

In Fig.~\ref{fig:AdobeComparison}, we show the enhanced results for the photo retouching application on the MIT-Adobe FiveK dataset~\cite{Adobe5K}. Each result is paired with its respective error map, displayed in the upper-left corner, with values ranging from 0 to 40. The results demonstrate that the proposed method yields visually superior results of 4D LUT and CoTF. Specifically, when the input images contain dominant colors (as in the first and second rows) or complex color variations (as in the third and last rows), 4D LUT and CoTF exhibit noticeable errors, whereas the proposed method consistently produces results with significantly lower errors.

In Table~\ref{table:adobe5K_4K}, we assess the applicability of the proposed method to high-resolution images. The proposed method again yields the best performance in terms of PSNR, SSIM, and $\Delta E_{ab}$, emphasizing the efficacy of the proposed method in handling high-resolution images.

We assess the proposed method against ten state-of-the-art approaches~\cite{HDRNet, CSRNet, 3DLUT, PPR10K, SepLUT, Adaint, RSFNet, HashLUT,4DLUT,CoTF} on the PPR10K dataset~\cite{PPR10K} as listed in Table~\ref{table:PPR10K}. Here, we additionally evaluate the model based on ResNet-34, called `Proposed-Res34'. `Proposed' uses ResNet-18, consistent with~\cite{PPR10K, SepLUT, Adaint}, ensuring a fair comparison, while `Proposed-Res34' incorporates a deeper backbone to assess scalability on large-scale and diverse datasets. As shown in Table~\ref{table:PPR10K}, `Proposed' outperforms other state-of-the-art methods in every metric. Moreover, the performance improvement of `Proposed-Res34' demonstrates the benefits of increased expressiveness for large-scale datasets.

Additionally, in Fig.~\ref{fig:PPR10KComparison}, we show qualitative results in which AdaInt and RSFNet exhibit noticeable pixel-level inaccuracies. In contrast, the proposed method yields results that closely match the ground-truth. These results confirm the efficacy of the proposed method in enhancing image quality through pigment-based representation.

\def\arraystretch{1.2}
\begin{table}[t]
    \centering
    \caption{Quantitative comparisons on the MIT-Adobe FiveK dataset~\cite{Adobe5K} for the tone mapping application. The best result is boldfaced.}
    \label{table:adobe5K_tone}
    \resizebox{0.63\columnwidth}{!}
    {
        \small
        \begin{tabular}{c | ccc} 
            \Xhline{3\arrayrulewidth}
            Methods & PSNR & SSIM & $\Delta E_{ab}$ \\
            \Xhline{3\arrayrulewidth}
            UPE~\cite{UPE} & 21.56 & 0.837 & 12.29\\
            DPE~\cite{DPE} & 22.93 & 0.894 & 11.09 \\
            HDRNet~\cite{HDRNet} & 24.52 & 0.915 & 8.14\\
            CSRNet~\cite{CSRNet} & 25.19 & 0.921 & 7.63\\
            3D LUT~\cite{3DLUT} & 25.29 & 0.920 & 7.55 \\
            SepLUT~\cite{SepLUT} & 25.43 & 0.922 & 7.43 \\

            AdaInt~\cite{Adaint} & 25.49 & 0.926 & 7.47\\
                4D LUT~\cite{4DLUT} & 25.21 & 0.931 & 7.38 \\
                CoTF~\cite{CoTF} & 25.47 & 0.938 & 7.37 \\
   Proposed & \textbf{25.71} & \textbf{0.940} & \textbf{7.15} \\
            \Xhline{3\arrayrulewidth}
        \end{tabular}
    }
\end{table}

\begin{table}[t]
    \centering
    \caption{The effect of $N$ and $L$ on the MIT-Adobe FiveK dataset~\cite{Adobe5K} for the photo retouching application.}
    \label{table:ablationNL}
    \resizebox{0.58\columnwidth}{!}
    {
        \small
        \begin{tabular}{cc | ccc} 
            \Xhline{3\arrayrulewidth}
            $N$ & $L$ & PSNR & SSIM & $\Delta E_{ab}$ \\
            \Xhline{3\arrayrulewidth}
                \multirow{4}{*}{16} & 8 & 25.61 & 0.936 & 7.36 \\ 
                 & 16 & 25.62 & 0.935 & 7.35 \\ 
               & 32 & 25.63 & 0.938 & 7.25 \\ 
               & 64 & 25.65 & 0.937 &  7.20 \\
               \hline
            
            \multirow{4}{*}{32} & 8 & 25.71  & 0.937 & 7.24  \\ 
                & 16 & 25.71 & \textbf{0.939} & 7.26  \\ 
                & 32 & 25.72 & 0.938  & 7.20 \\ 
                & 64 & 25.70  & 0.936 & 7.18 \\

                \hline

                \multirow{4}{*}{64} & 8 & 25.68 & \textbf{0.939} & 7.29 \\ 
             & 16 & 25.72 & \textbf{0.939} & 7.19 \\ 
             & 32 & \textbf{25.82} & \textbf{0.939} & \textbf{7.15} \\ 
             & 64 & 25.63 & 0.937 & 7.20 \\
                \hline

                \multirow{4}{*}{128} & 8 & 25.67 & \textbf{0.939} & 7.26 \\ 
             & 16 & 25.66 & \textbf{0.938} & 7.24  \\ 
             & 32 & 25.71  & \textbf{0.939} & 7.20  \\ 
             & 64 & 25.69 & \textbf{0.939} & 7.20 \\
            
            \Xhline{3\arrayrulewidth}
        \end{tabular}
    }
\end{table}

For subjective assessment, we conducted a user study. Specifically, we randomly selected 50 images from the test set of the MIT-Adobe FiveK dataset~\cite{Adobe5K} and presented the enhanced results of 4D LUT~\cite{4DLUT}, CoTF~\cite{CoTF} to 10 participants in random order. Then, participants were asked to choose the most visually pleasing and similar result to the corresponding expert-retouched image. In total, 500 votes (50 images $\times$ 10 participants) were cast. The proposed method won more votes: it was preferred in 39.2\%, while CoTF and 4D LUT were preferred in 26.8\% and 34.2\%, respectively.

\subsubsection{Tone mapping}
\begin{figure*}[]
    \centering
    \subfigure{\includegraphics[width=4cm]{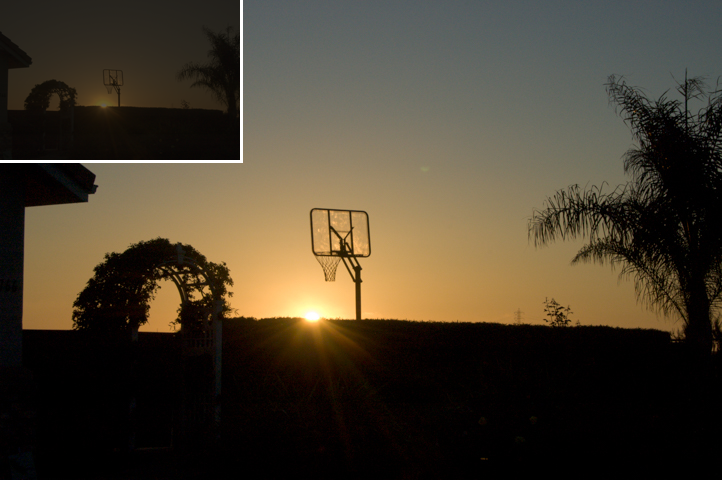}}  \hspace{0.02cm}
    \subfigure{\includegraphics[width=4cm]{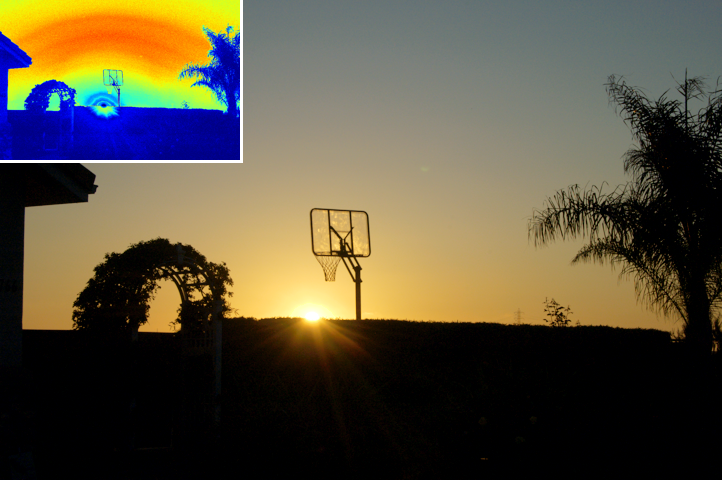}} \hspace{0.02cm}
    \subfigure{\includegraphics[width=4cm]{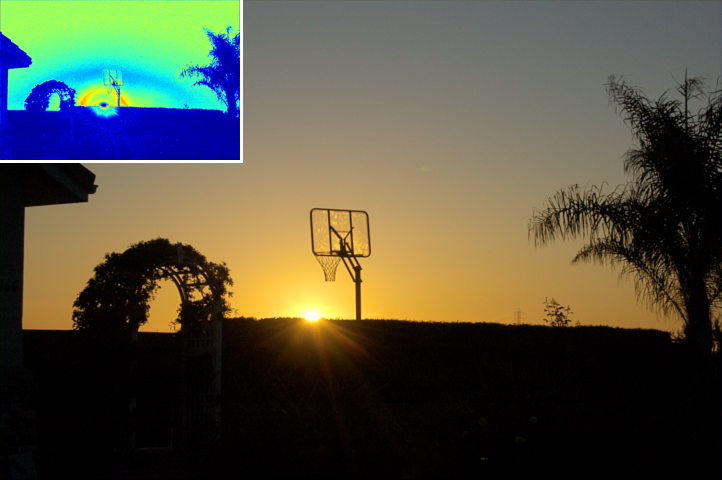}} \hspace{0.02cm}
    \subfigure{\includegraphics[width=4cm]{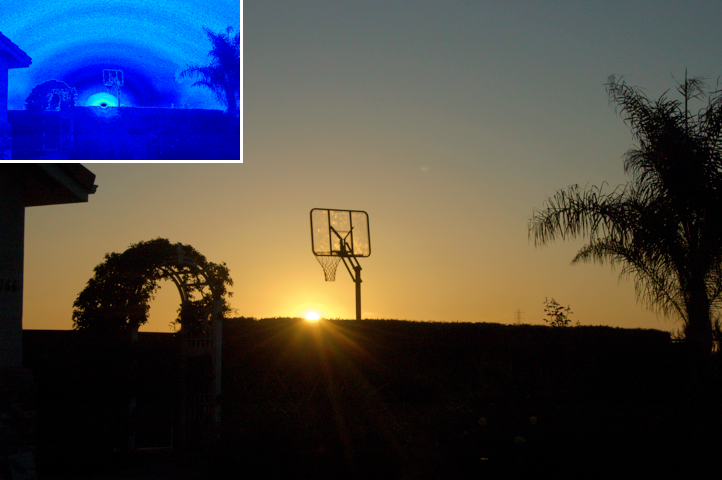}} \hspace{0.01cm}
    \subfigure{\includegraphics[width=0.258cm]{Figures/ppr10k/colorbar.png}} \\
    \setcounter{subfigure}{0}
    \subfigure[GT]{\includegraphics[width=4cm]{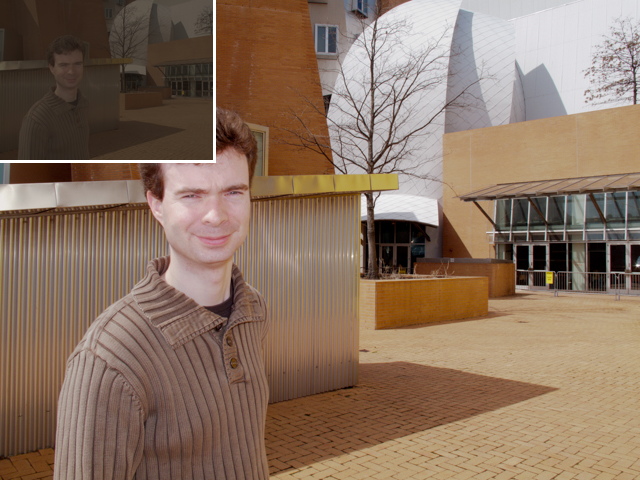}}  \hspace{0.02cm}
    \subfigure[AdaInt]{\includegraphics[width=4cm]{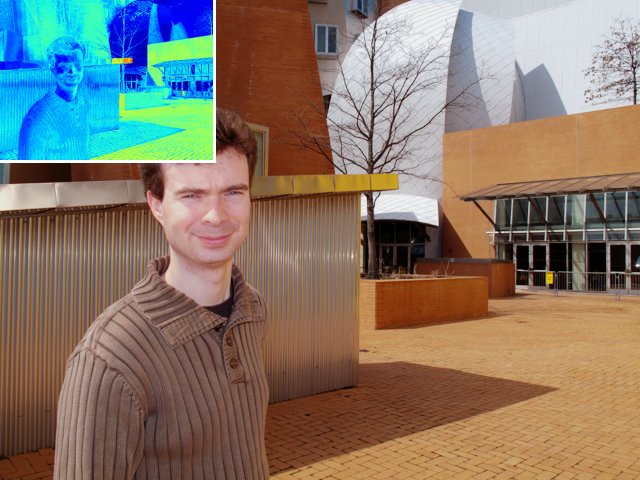}} \hspace{0.02cm}
    \subfigure[CoTF]{\includegraphics[width=4cm]{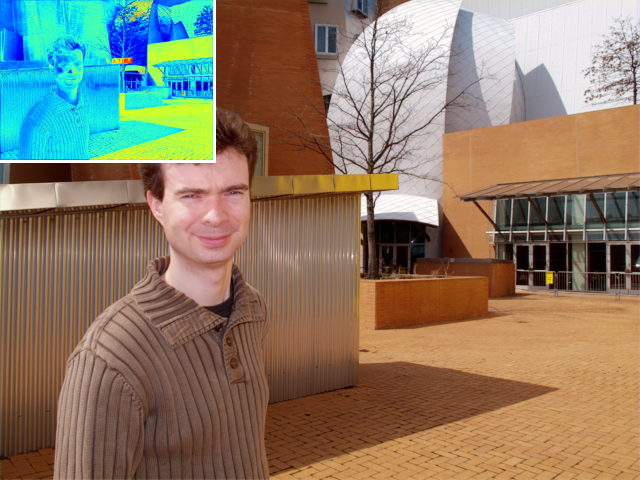}} \hspace{0.02cm}
    \subfigure[Proposed]{\includegraphics[width=4cm]{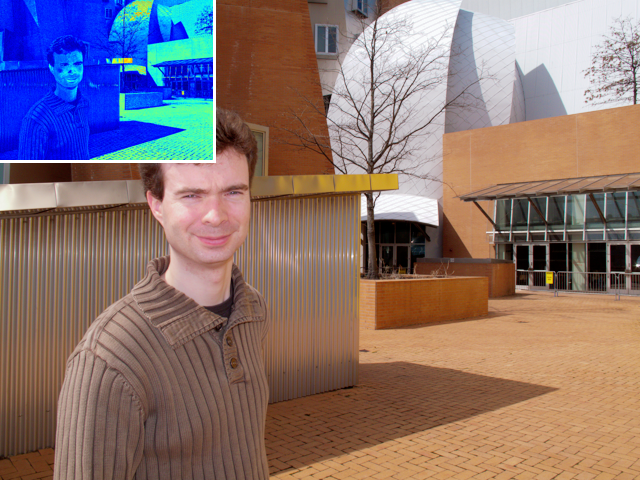}} \hspace{0.01cm}
    \subfigure{\includegraphics[width=0.258cm]{Figures/ppr10k/colorbar.png}}
    \vspace{-0.1cm}
    \caption{Qualitative comparisons on the MIT-Adobe FiveK dataset~\cite{Adobe5K} for tone mapping: (a) shows GT (retouched by expert C) with its corresponding input image. The input images are represented as 8-bit sRGB format images, achieved by normalizing the 16-bit CIE XYZ values and applying the transformation matrix described in \cite{xyz2rgb}. (b), (c), and (d) show the resultant images and their corresponding error maps obtained by AdaInt~\cite{Adaint}, CoTF~\cite{CoTF}, and the proposed method, respectively.}
\label{fig:toneComparison}
\end{figure*}

In Table~\ref{table:adobe5K_tone}, we compare the proposed method with nine state-of-the-art methods~\cite{UPE,DPE,HDRNet,CSRNet,3DLUT,Adaint,SepLUT,4DLUT,CoTF}. We confirm that the proposed method achieves superior performance compared to the other methods in all metrics. Fig.~\ref{fig:toneComparison} illustrates the qualitative results for the MIT-Adobe FiveK dataset~\cite{Adobe5K}, in which AdaInt and CoTF suffer from color tone discrepancies while the proposed method yields results with substantially fewer errors.

\begin{table*}[t]
    \centering
        \caption{Component and structure variations of the proposed method on the MIT-Adobe FiveK dataset~\cite{Adobe5K} for the photo retouching application. The best result is boldfaced.}
    \resizebox{1.7\columnwidth}{!}
    {
    \begin{tabular}{c|c|cccc|cccc}
    \Xhline{2\arrayrulewidth}
    Structures & Exp. No. & $\mathbf{W}$ & Pigment reprojection & $\mathbf{U}$ & Pigment blending & PSNR & SSIM  & $\Delta E_{ab}$ & Runtime \\ \hline
    \multirow{7}{*}{Variations} & (a) & \checkmark &  & & \checkmark & 24.07 & 0.914      & 8.51 & 1.32ms \\
    & (b) & & \checkmark & & \checkmark & 25.58 & 0.938 & 7.27 & 1.40ms \\
    & (c) & & & \checkmark & \checkmark & 25.38 & 0.932 & 7.33 & 1.33ms\\
    & (d) &  & \checkmark & \checkmark & \checkmark & 25.61 &  \textbf{0.939} & 7.25 & 1.42ms \\
    & (e) & \checkmark & & \checkmark & \checkmark & 25.54 & 0.936 & 7.26 & 1.35ms \\
    & (f) & \checkmark & \checkmark &  & \checkmark & 25.64 & \textbf{0.939} & 7.18 & 1.41ms \\ 
    & (g) & \checkmark & \checkmark & \checkmark &  & 25.70 &  \textbf{0.939} & 7.19 & 1.33ms \\ \hline
    Proposed & (h) & \checkmark & \checkmark & \checkmark & \checkmark & \textbf{25.82} & \textbf{0.939} & \textbf{7.15} & 1.43ms \\
    \Xhline{2\arrayrulewidth}
    \end{tabular}
    }
\label{table:ablation}
\end{table*}

\begin{table}[t!]
    \centering
    \caption{The variation of the backbone network on the MIT-Adobe FiveK dataset~\cite{Adobe5K} for the photo retouching application. The best result is boldfaced.}
    \label{table:ablationBackbone}
    \resizebox{0.8\columnwidth}{!}
    {
        \small
        \begin{tabular}{c|c | ccc} 
            \Xhline{3\arrayrulewidth}
            Backbones & Methods & PSNR & SSIM & $\Delta E_{ab}$ \\
            \Xhline{3\arrayrulewidth}
            \multirow{4}{*}{5-layer} & 3D LUT~\cite{3DLUT} & 25.29 & 0.920 & 7.55 \\
            & SepLUT~\cite{SepLUT} & 25.47 & 0.921 & 7.54 \\
            & AdaInt~\cite{Adaint} & 25.49 & 0.926 & 7.47 \\
            & Proposed & \textbf{25.82} & \textbf{0.939} & \textbf{7.15} \\
            \hline
            \multirow{4}{*}{ResNet-18} & 3D LUT~\cite{3DLUT} & 25.23 & 0.930 & 7.26 \\
            & SepLUT~\cite{SepLUT} & 25.30 & 0.934 & 7.63 \\
            & AdaInt~\cite{Adaint} & 25.24 & 0.933 & 7.61 \\
            & Proposed & \textbf{25.95} & \textbf{0.941} & \textbf{7.06} \\
            \hline
            ResNet-34 & Proposed & \textbf{25.97} & \textbf{0.941} & \textbf{7.02} \\
            \Xhline{3\arrayrulewidth}
        \end{tabular}
    }
\end{table}

\begin{figure*}[t]
    \centering
    \subfigure{\includegraphics[width=4cm]{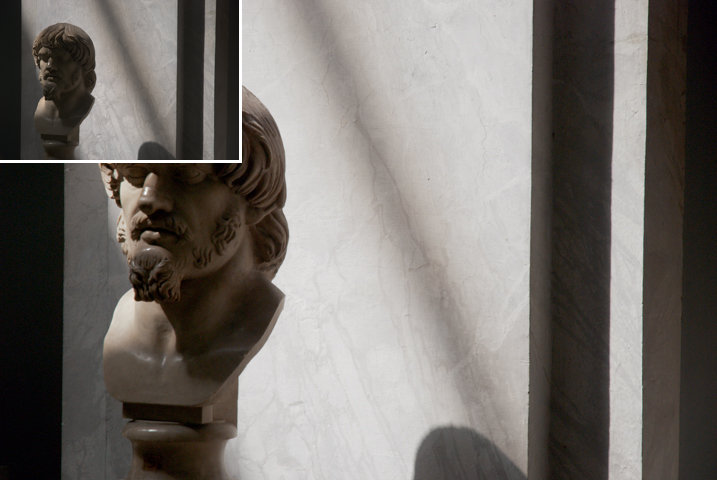}}  \hspace{0.02cm}
    \subfigure{\includegraphics[width=4cm]{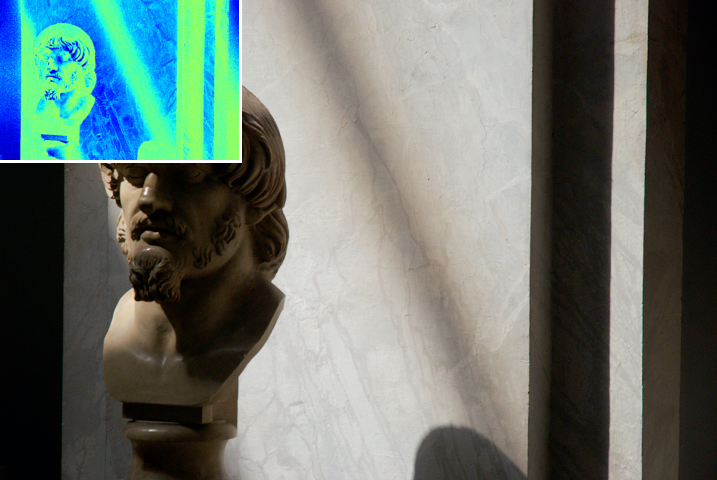}} \hspace{0.02cm}
    \subfigure{\includegraphics[width=4cm]{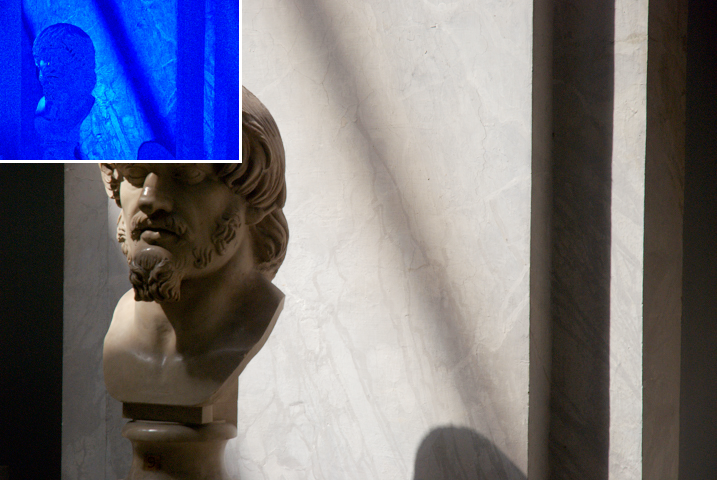}} \hspace{0.02cm}
    \subfigure{\includegraphics[width=4cm]{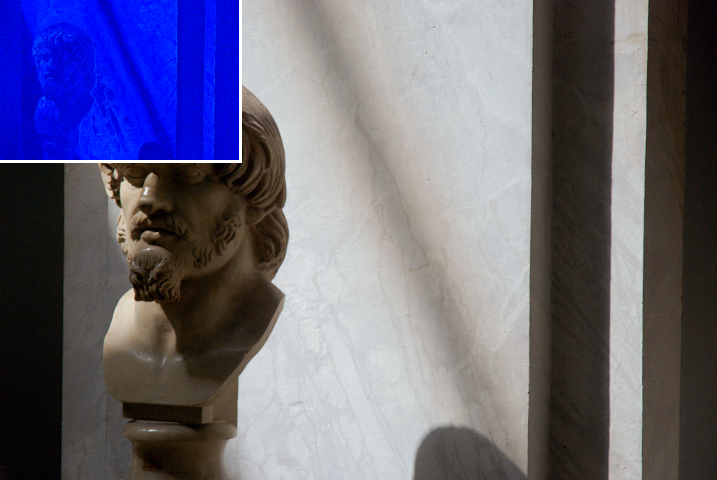}} \hspace{0.01cm}
    \subfigure{\includegraphics[width=0.258cm]{Figures/ppr10k/colorbar.png}} \\

    \setcounter{subfigure}{0}
    \subfigure[GT]{\includegraphics[width=4cm]{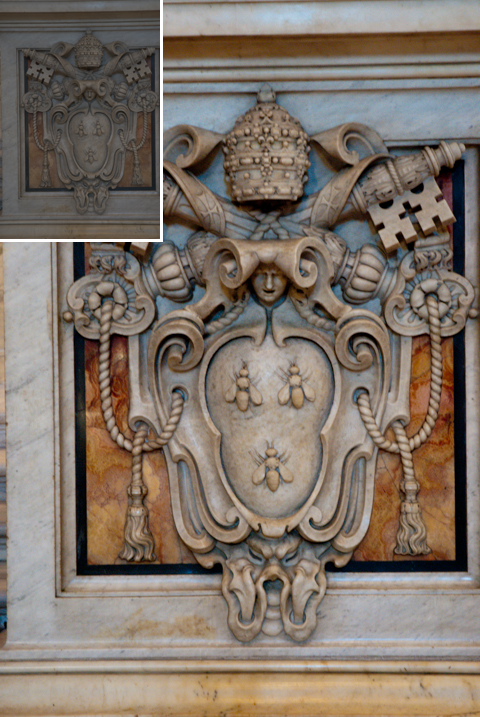}}  \hspace{0.02cm}
    \subfigure[5-layer]{\includegraphics[width=4cm]{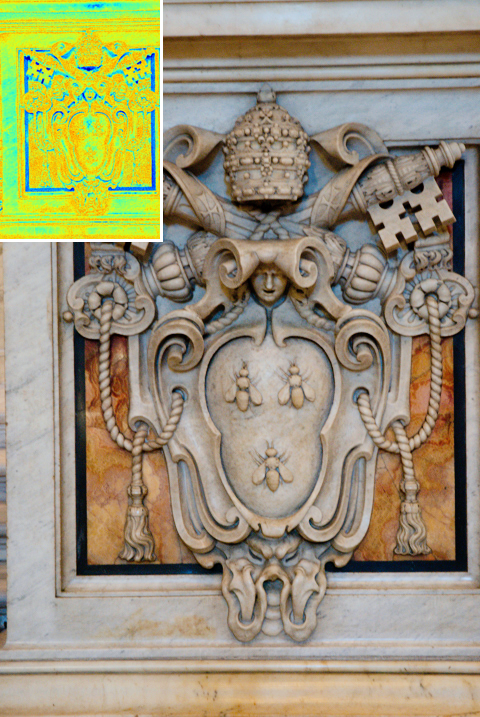}} \hspace{0.02cm}
    \subfigure[ResNet-18]{\includegraphics[width=4cm]{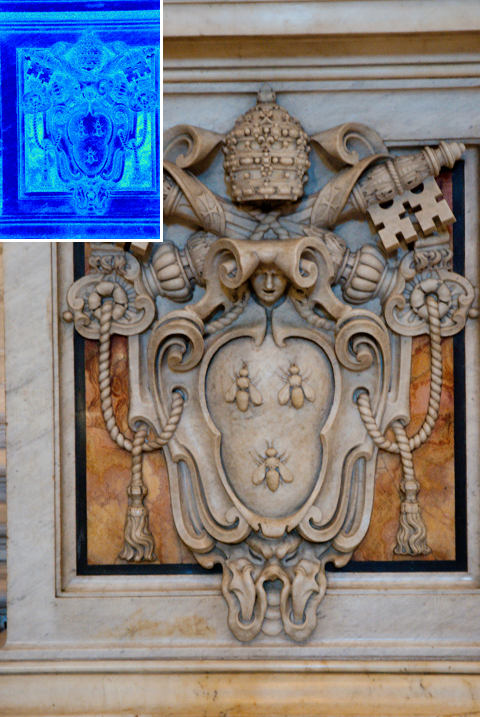}} \hspace{0.02cm}
    \subfigure[ResNet-34]{\includegraphics[width=4cm]{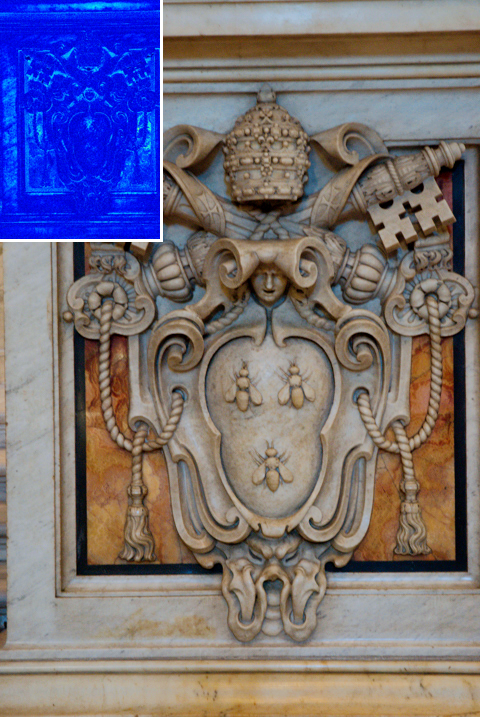}} \hspace{0.01cm}
    \subfigure{\includegraphics[width=0.258cm]{Figures/ppr10k/colorbar_long.png}}
    \vspace{-0.1cm}
    \caption{Qualitative comparisons on the MIT-Adobe FiveK dataset~\cite{Adobe5K} using different backbone networks for image retouching: (a) shows GT (retouched by expert C) with its corresponding input image. (b), (c), and (d) show the resultant images and their corresponding error maps using 5-layer, ResNet-18, and ResNet-34 backbones, respectively.}
\label{fig:backboneTest}
\end{figure*}

\begin{figure*}[h!]
    \centering
    \includegraphics[width=\textwidth]{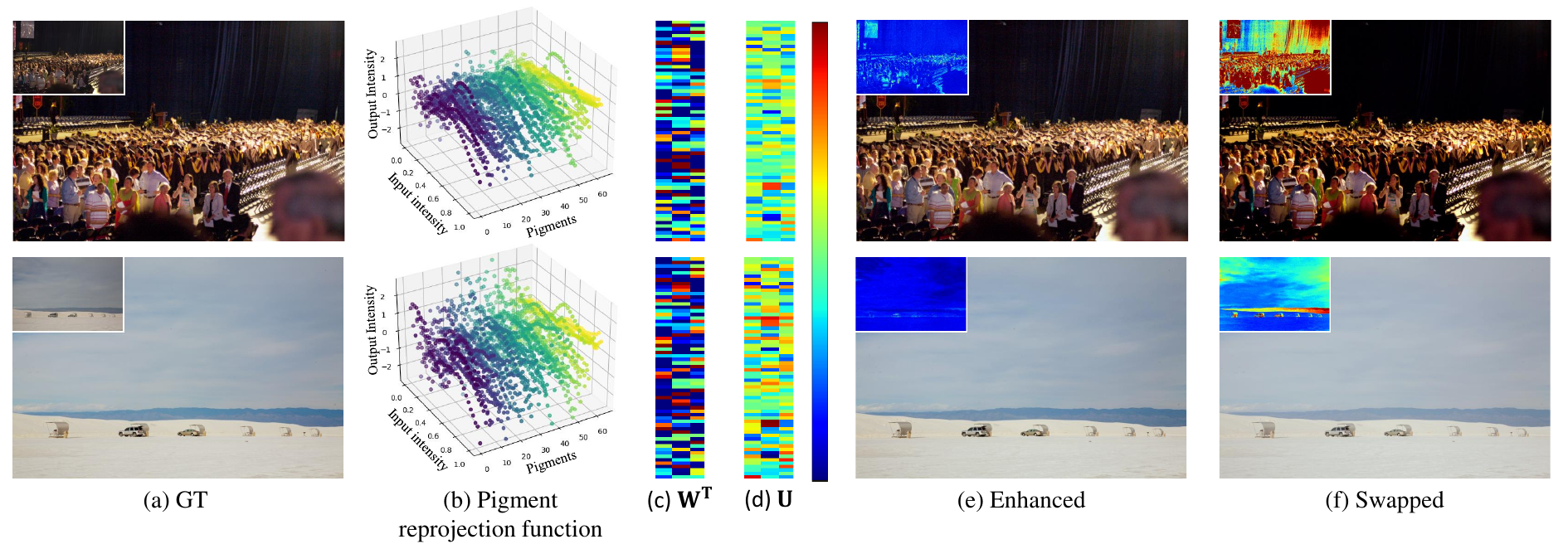}
    \vspace{-0.3cm}
    \caption
    {
        Visualization of the proposed pigment representation: (a) is GT with its corresponding input image, (b) is the pigment reprojection function, (c) is the pigment expansion weight $\mathbf{W}$, (d) is the reconstruction weight $\mathbf{U}$, and (e) is the enhanced image. For (b) the pigment reprojection function, it is represented as a set of tweaked points as described in Eq. (\ref{eq:baro}). The values of (c) the pigment expansion weight and (d) the reconstruction weight are represented by color, with higher values shown in red and lower values in blue. Also, (f) shows the enhanced image when the pigment-related components (b)–(d) are swapped between the upper and lower input images in (a).
    }
    \label{fig:visualization}
\end{figure*}

\begin{figure}
    \centering
    \renewcommand{\arraystretch}{1.2}
    \begin{tabular}{@{}c@{\hspace{0.1cm}}c@{\hspace{0.15cm}}c@{}} 
        \raisebox{0.7cm}{\rotatebox{90}{\scriptsize $1 \times 10^{-3}$}} &
        \includegraphics[width=3.5cm]{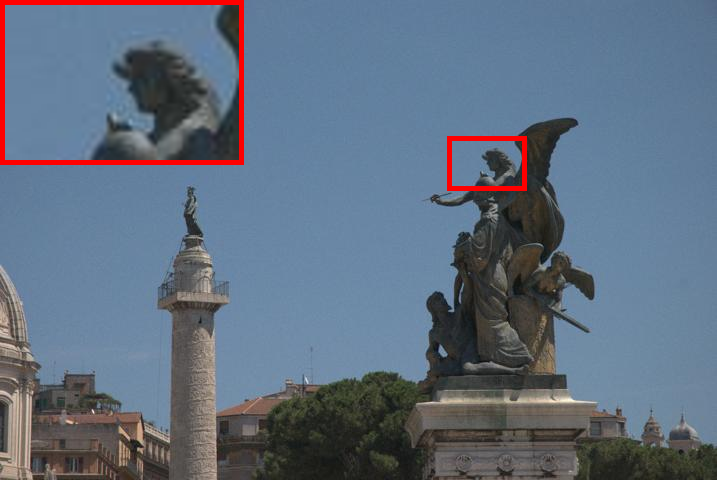} &
        \includegraphics[width=3.5cm]{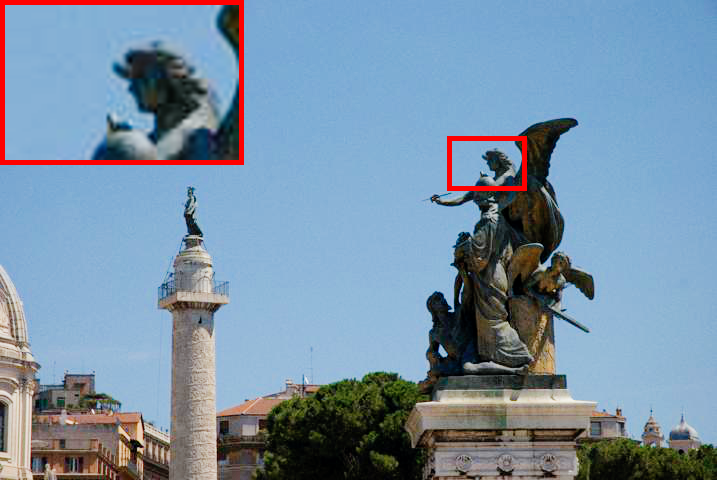} \\
        
        \raisebox{0.7cm}{\rotatebox{90}{\scriptsize $5 \times 10^{-3}$}} &
        \includegraphics[width=3.5cm]{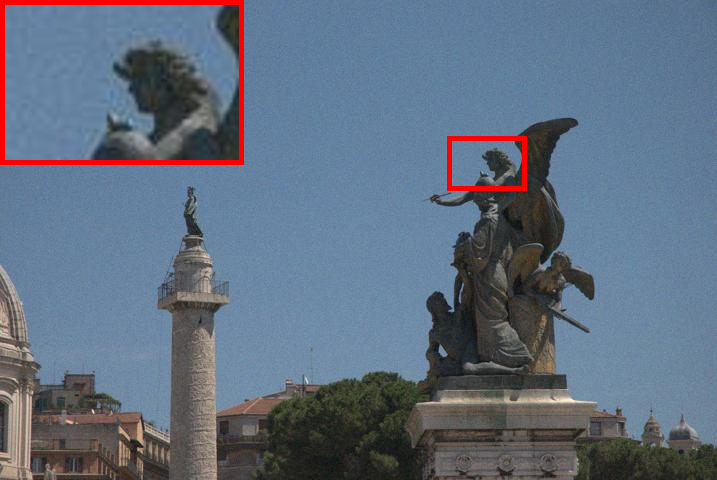} &
        \includegraphics[width=3.5cm]{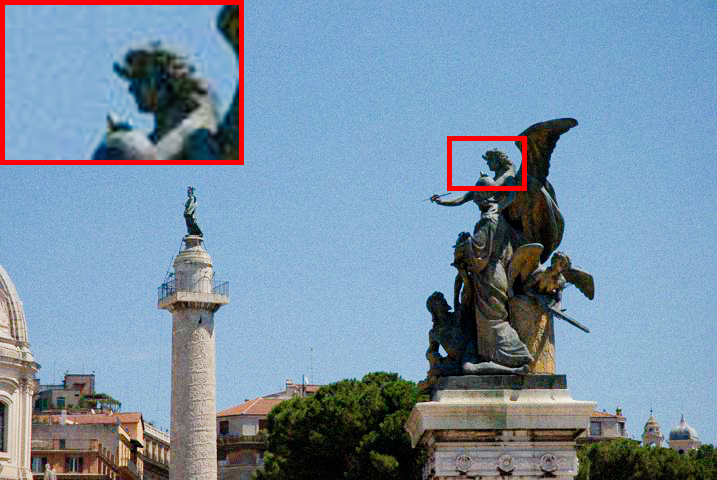} \\
        
        \raisebox{0.65cm}{\rotatebox{90}{\scriptsize $2.5 \times 10^{-2}$}} &
        \includegraphics[width=3.5cm]{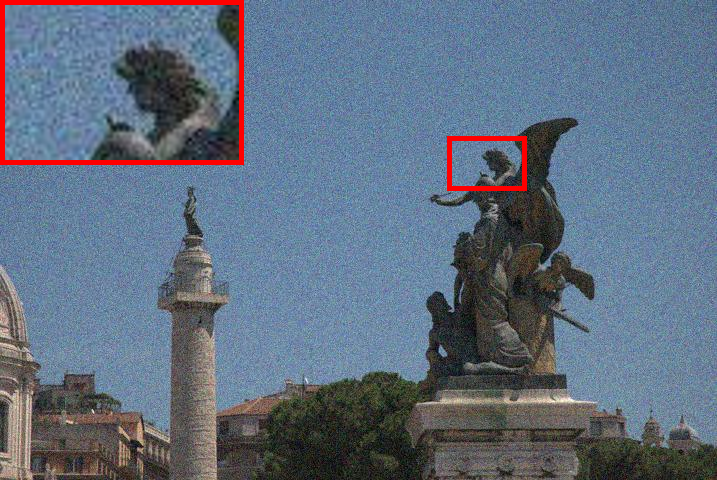} &
        \includegraphics[width=3.5cm]{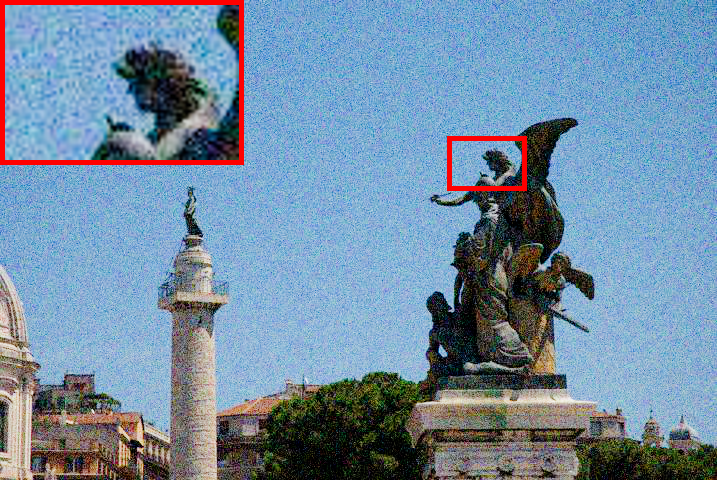} \\
        & \footnotesize{(a) Noisy input} & \footnotesize{(b) Enhanced}
    \end{tabular}
    \caption{Limitation examples of the proposed method. (a) shows input images corrupted by Gaussian noise with $\sigma^2 = 1 \times 10^{-3}$, $5 \times 10^{-3}$, and $2.5 \times 10^{-2}$, respectively; (b) shows the corresponding enhanced outputs.}
    \label{fig:limitations}
\end{figure}

\subsection{Ablation Studies and analyses}
\label{ssec:ablation}
\subsubsection{The effect of $N$ and $L$}
Table~\ref{table:ablationNL} compares photo retouching performance according to $N$ and $L$, where $N$ denotes the number of pigments, and $L$ is the number of reprojection offsets for each pigment. It is observed that increasing $N$ until it reaches $64$ results in performance improvement in every metric. On the other hand, when $N=128$, the PSNR scores decrease while SSIM and $\Delta E_{ab}$ performance does not improve. Similarly, as $L$ increases to $32$, performance improves, and when $L=64$, it degrades or maintains. These results indicate that adopting a sufficient number of pigments and offsets enhances expressiveness and leads to reliable performance. Thus, we set $N$ and $L$ to 64 and 32, respectively.

\subsubsection{Component analysis}
Table~\ref{table:ablation} compares several ablated methods to analyze the efficacy of the proposed method and its components. The evaluation is conducted on the photo retouching task using the MIT-Adobe FiveK dataset~\cite{Adobe5K}. To show the efficacy of the pigment expansion and RGB reconstruction in Eq. (\ref{eq:f}) and Eq. (\ref{eq:base_point4}), we replace $\mathbf{W}$ and $\mathbf{U}$ with trainable parameters that are determined during training regardless of the input image.

\subsubsection{The effect of the backbone network}
Table~\ref{table:ablationBackbone} illustrates the efficacy of the backbone network for 3D LUT~\cite{3DLUT}, SepLUT~\cite{SepLUT}, AdaInt~\cite{Adaint}, and the proposed method. We conduct photo retouching experiments using the 5-layer backbone of \cite{3DLUT} and ResNet-18. ResNet-34 is only applied to the proposed method due to the compatibility issue. While the conventional methods fail to enhance performance when replacing the backbone with a heavier model, the proposed method successfully boosts performance and performs the best for all backbones. Fig.~\ref{fig:backboneTest} illustrates that the qualitative performance of the proposed method improves as heavier backbone networks are employed. This experiment indicates that the proposed method can be applied across various backbone networks, with performance improvements achievable by adopting heavier backbone networks.

In settings (a)-(c), we develop the models without two components. In the remaining settings, only one component is replaced or removed. These results show that all proposed components contribute significantly to enhancing performance, with only a negligible increase in runtime (up to 0.11 ms). Especially, the gap between settings (e) and (h) demonstrates that the proposed pigment reprojection contributes the most to performance improvement. The comparison between settings (d) and (h) indicates that pigment expansion provides the second-largest performance gain, which enables the acquisition of scene-adaptive characteristics.

\subsubsection{Analysis of the proposed pigment representation}
\label{ssec:visualization}
To verify the input-adaptive operation of the proposed method, we visualize enhancement results for two different images and their pigment-related components~(\ie, pigment reprojection functions, pigment expansion weights, and reconstruction weights) in Fig.~\ref{fig:visualization}. In Figs.~\ref{fig:visualization}(b)-(d), these components are adaptively estimated based on inputs. These distinct estimations enable expressive and high-quality enhancements as in Fig.~\ref{fig:visualization}(e).

Also, to validate the adaptability of the pigment representation, we performed an additional experiment, as shown in Fig.~\ref{fig:visualization}(f). In this experiment, the pigment-related components were exchanged between the upper and lower images in Fig.~\ref{fig:visualization}. These results indicate that when the pigment representation does not align with the input content, the enhancement quality is degraded. Specifically, comparing Fig.~\ref{fig:visualization}(f) to Fig.~\ref{fig:visualization}(e) reveals a significant degradation in quality. These results clearly demonstrate that our method effectively adapts transformations to each image’s content, leading to improved performance.

\section{Limitations and Future Work}
\label{sec:limitations}
As demonstrated in Section~\ref{sec:exp}, the proposed method achieves high performance both qualitatively and quantitatively. However, since it is a global transformation-based approach, it inherently struggles to capture local information. Consequently, when input images contain noise or localized distortions, the enhancement quality may degrade due to the lack of spatial adaptiveness. As illustrated in Fig.~\ref{fig:limitations}, the proposed method suffers from performance degradation under three different levels of Gaussian noise, and the degradation becomes more pronounced as the noise level increases.

To address this limitation, we plan to explore a hybrid approach that integrates dense mapping-based techniques into the proposed method. By introducing local adaptiveness while preserving the efficiency of the global transformation, we aim to enhance robustness against noise and spatial inconsistencies.

Additionally, we plan to extend the proposed pigment representation-based enhancement method to other challenging image enhancement tasks, such as low-light~\cite{LL1,LL2} and underwater image enhancement~\cite{UE}. These domains present unique challenges, including extreme contrast variations and color distortions, where our method can be further optimized and adapted to achieve superior performance.

\section{Conclusions}
\label{sec:conclusions}

We have introduced a novel image enhancement method based on pigment representation, which transforms RGB colors into a sophisticated high-dimensional feature space called \textit{pigment} and refines them for image enhancement. This innovative pigment representation offers two main innovations. First, the customized conversion of RGB colors to pigments is designed to consider input content, ensuring adaptability to diverse image content. Second, the abundance of pigments enhances expressiveness, leading to superior image enhancement performance. The proposed method is structured into five stages. Initially, an image is input into the visual encoder to obtain parameters such as pigment expansion weights, reprojection offsets, and RGB reconstruction weights. Subsequently, pigment expansion translates RGB colors into pigments through a weighted sum calculation, utilizing adaptive weight values from the visual encoder to ensure adaptability to the content. In the following stage, pigment reprojection individually reprojects each pigment using the corresponding pigment reprojection function determined by the earlier acquired reprojection offsets to refine its positioning within the pigment feature space. To enhance pigment correlations and capture complex color dynamics, a pigment blending stage is implemented, employing two convolution layers. In the final stage, RGB reconstruction generates reconstructed images by computing the weighted sum of blended pigments, employing RGB reconstruction weights from the visual encoder. Comprehensive experimental results highlight the superior performance of the proposed method compared to state-of-the-art techniques in various image enhancement tasks, including image retouching and tone mapping. Notably, the method achieves this while maintaining a relatively low level of computational complexity and a compact model size.
 
\bibliographystyle{IEEEtran}
\bibliography{main_arxiv}

\end{document}